\pdfoutput=1

\documentclass[camera,letterpaper,nomarginnotes,nonarrowgutter]{jpaper}

\PassOptionsToPackage{hyphens}{url}
\usepackage[bookmarks=true,breaklinks=true,colorlinks,linkcolor=black,citecolor=black,urlcolor=blue]{hyperref}

\usepackage[normalem]{ulem}
\usepackage{subcaption}
\usepackage{clipboard}
\usepackage{soul}
\usepackage[dvipsnames]{xcolor}
\usepackage{tikz}
\usepackage{flushend}
\usepackage[noadjust]{cite}
\usepackage{xargs} 
\usepackage{mathtools}
\usepackage{enumitem}
\usepackage{booktabs}
\usepackage{multirow}
\usepackage[many]{tcolorbox}
\usepackage{longfbox}
\usepackage{listings}
\usepackage{algorithmic}
\usepackage{graphicx}
\usepackage{textcomp}
\usepackage{pifont}
\usepackage[acronym,nonumberlist,nowarn]{glossaries}
    \glsdisablehyper
     \newacronym{ADI}{ADI}{Alternating Direction Implicit}
\newacronym{AGU}{AGU}{Address Generation Unit}
\newacronym[longplural={Access History Tables}]{AHT}{AHT}{Access History Table}
\newacronym{AHT-R}{AHT-R}{Access History Table - Read}
\newacronym{AHT-W}{AHT-W}{Access History Table - Write-Back}
\newacronym{DMA}{DMA}{direct memory access}

\newacronym{AIP}{AIP}{Access Interval Predictor}
\newacronym{AL}{AL}{Added Latency for column accesses}
\newacronym{ASIC}{ASIC}{Application Specific Integrated Circuit}
\newacronym{AVX}{AVX}{Advanced Vector Extensions}
\newacronym[longplural={Basic Block Vectors}]{BBV}{BBV}{Basic Block Vector}
\newacronym{BHT}{BHT}{Branch History Table}
\newacronym{HBM}{HBM}{Hybrid Bandwidth Memory}
\newacronym{DDR4}{DDR4}{Double-Data Rate 4}
\newacronym{MIMD}{MIMD}{multiple-instruction multiple-data}
\newacronym{SIMT}{SIMT}{single-instruction multiple-threads}
\newacronym{SLP}{SALP}{subarray-level parallelism}
\newacronym{BLP}{BLP}{bank-level parallelism}

\newacronym{NN}{NN}{neural network}
\newacronym{BTB}{BTB}{Branch Target Buffer}
\newacronym{CAS}{CAS}{Column Address Strobe}
\newacronym{AMAT}{AMAT}{Average Memory Access Time}
\newacronym{CCD}{CCD}{Column to Column Delay}
\newacronym{AI}{AI}{Arithmetic Intensity}
\newacronym[longplural={Columnar Database Systems}]{CDBS}{CDBS}{Columnar Database System}
\newacronym[longplural={Chip Multiprocessors}]{CMP}{CMP}{Chip Multiprocessor}
\newacronym{CMT}{CMT}{Chip Multithreading}
\newacronym[longplural={Coarse-Grain Reconfigurable Arrays}]{CGRA}{CGRA}{Coarse-Grain Reconfigurable Array}
\newacronym{C-RAM}{C-RAM}{Computational-RAM}
\newacronym{CWD}{CWD}{Column Write Delay}
\newacronym{DBI}{DBI}{Dynamic Binary Instrumentation}
\newacronym[longplural={Database Management Systems}]{DBMS}{DBMS}{Database Management System}
\newacronym{DDR}{DDR}{Double Data Rate}
\newacronym{DEWP}{DEWP}{Dead Line and Early Write-Back Predictor}
\newacronym{DRAM}{DRAM}{Dynamic Random Access Memory}
\newacronym{DSBP}{DSBP}{Dead Sub-Block Predictor}
\newacronym{DSM}{DSM}{Decomposed Storage Manage}
\newacronym{FAW}{FAW}{Four row Activation Window}
\newacronym{FP}{FP}{Floating-Point}
\newacronym{EDP}{EDP}{Energy-Delay Product}
\newacronym{FCFS}{FCFS}{First-Come First-Serve}
\newacronym{FIFO}{FIFO}{Fist In, First Out}
\newacronym{FSM}{FSM}{finite state machine}
\newacronym{FPGA}{FPGA}{Field-Programmable Gate Array}
\newacronym[longplural={Functional Units}]{FU}{FU}{Functional Unit}
\newacronym{GAg}{GAg}{Global Adaptive branch prediction using one Global PHT}
\newacronym{GAs}{GAs}{Global Adaptive branch prediction using per-Set PHT}
\newacronym{GCC}{GCC}{GNU Compiler Collection}
\newacronym{GEMS}{GEMS}{General Execution-driven Multiprocessor Simulator}
\newacronym[longplural={General Purpose Processors}]{GPP}{GPP}{General Purpose Processor}
\newacronym{HMC}{HMC}{Hybrid Memory Cube}
\newacronym{HIVE}{HIVE}{HMC Instruction Vector Extensions}
\newacronym{IATAC}{IATAC}{Inter-Access Time per Access Count}
\newacronym{ILP}{ILP}{Instruction Level Parallelism}
\newacronym{IPC}{IPC}{Instructions per Cycle}
\newacronym{ISA}{ISA}{Instruction Set Architecture}
\newacronym{IRAM}{IRAM}{Intelligent RAM}
\newacronym{KIPS}{KIPS}{Kilo Instructions per Second}
\newacronym{LDS}{LDS}{Linked Data Structure}
\newacronym{LFSR}{LFSR}{Linear Feedback Shift Register}
\newacronym{LFMR}{LFMR}{Last-to-First Miss-Ratio}
\newacronym{MLP}{MLP}{Memory-Level Parallelism}

\newacronym{LLC}{LLC}{last-level cache}
\newacronym{LRU}{LRU}{Least Recently Used}
\newacronym{LSU}{LSU}{Load Store Unit}
\newacronym{LTP}{LTP}{Last-Touch Predictor}
\newacronym{LvP}{LvP}{Live-time Predictor}
\newacronym{LWP}{LWP}{Last Write Predictor}
\newacronym{MARSS}{MARSS}{Micro Architectural and System Simulator}
\newacronym{McPAT}{McPAT}{Multi-core Power, Area, and Timing}
\newacronym{MIPS}{MIPS}{Microprocessor without Interlocked Pipeline Stages}
\newacronym{MOB}{MOB}{Memory Order Buffer}
\newacronym{MMU}{MMU}{memory management unit}
\newacronym{PuD}{PUD}{processing-using-DRAM}
\newacronym{MPKI}{MPKI}{Misses per Kilo-Instruction}
\newacronym{MSHR}{MSHR}{Miss-Status Handling Registers}
\newacronym{NAS}{NAS}{Numerical Aerodynamic Simulation}
\newacronym{NDP}{NDP}{Near-Data Processing}
\newacronym{NMP}{NMP}{Near Memory Processor}
\newacronym{NoC}{NoC}{Network-on-Chip}
\newacronym{NPB}{NPB}{NAS Parallel Benchmark}
\newacronym{NUCA}{NUCA}{Non-Uniform Cache Architecture}
\newacronym{NUMA}{NUMA}{Non-Uniform Memory Access}
\newacronym{OoO}{OoO}{Out-of-Order}
\newacronym{OpenMP}{OpenMP}{Open Multi-Processing}
\newacronym{OS}{OS}{Operating System}
\newacronym{PAg}{PAg}{Per-address Adaptive branch prediction using one Global PHT}
\newacronym{PAs}{PAs}{Per-address Adaptive branch prediction using per-Set PHT}
\newacronym{PC}{PC}{Program Counter}
\newacronym[longplural={Pointer-Chasing Engines}]{PCE}{PCE}{Pointer-Chasing Engine}
\newacronym{PCM}{PCM}{Performance Counter Monitor}
\newacronym{PIM}{PIM}{processing-in-memory}
\newacronym[longplural={Pattern History Tables}]{PHT}{PHT}{Pattern History Table}
\newacronym{RAPL}{RAPL}{Running Average Power Limit}
\newacronym{GEMM}{GEMM}{general matrix-matrix multiplication}
\newacronym{RAS}{RAS}{Row Address Strobe}
\newacronym{RAT}{RAT}{Registers Alias Table}
\newacronym{RC}{RC}{Row Cycle}
\newacronym{RCD}{RCD}{RAS to CAS Delay}
\newacronym[longplural={Row-based Database Systems}]{RDBS}{RDBS}{Row-based Database System}
\newacronym{ROB}{ROB}{Reorder Buffer}
\newacronym{RRD}{RRD}{Row to Row activation Delay}
\newacronym{DNN}{DNN}{Deep Neural Network}
\newacronym{ANN}{ANN}{Artificial Neural Network}
\newacronym{PnM}{PNM}{processing-near-memory}
\newacronym{PuM}{PUM}{processing-using-memory}
\newacronym{FFD}{FFD}{first-fit-decreasing}
\newacronym{HFF}{HFF}{helper flip-flop}
\newacronym{OBPS}{OBPS}{one-bit per-subarray}
\newacronym{ABOS}{ABOS}{all-bits in one-subarray}
\newacronym{ABPS}{APBS}{all-bits per-subarray}
\newacronym{GEMV}{GEMV}{general matrix-vector product}

\newacronym{RBR}{RBR}{redundant binary representation}
\newacronym{RP}{RP}{Row Precharge}
\newacronym{RTL}{RTL}{Register Transfer Level}
\newacronym{RTP}{RTP}{Read To Precharge}
\newacronym{RVU}{RVU}{Reconfigurable Vector Unit}
\newacronym{SDP}{SDP}{Skewed Dead-Block Predictor}
\newacronym{SESC}{SESC}{Superescalar Simulator}
\newacronym{SSE}{SSE}{Streaming SIMD Extensions}
\newacronym{SFP}{SFP}{Spatial Footprint Predictor}
\newacronym{SiNUCA}{SiNUCA}{Simulator of Non-Uniform Cache Architectures}
\newacronym{SMT}{SMT}{Simultaneous Multi-Threading}
\newacronym{SIMD}{SIMD}{single-instruction multiple-data}
\newacronym{LUT}{LUT}{lookup table}

\newacronym{SPEC}{SPEC}{Standard Performance Evaluation Corporation}
\newacronym{SoC}{SoC}{System-on-Chip}
\newacronym{SPP}{SPP}{Spatial Pattern Predictor}
\newacronym{SRAM}{SRAM}{Static Random Access Memory}
\newacronym{SSOR}{SSOR}{Symmetric Successive Over-Relaxation}
\newacronym{SSV}{SSV}{Search Set Vector}
\newacronym{TLB}{TLB}{Translation Look-aside Buffer}
\newacronym{TLP}{TLP}{Thread Level Parallelism}
\newacronym[longplural={Through-Silicon Vias}]{TSV}{TSV}{Through-Silicon Via}
\newacronym{VWQ}{VWQ}{Virtual Write Queue}
\newacronym{WTR}{WTR}{Write Recovery time}
\newacronym{WR}{WR}{Write To Read delay time}
\newacronym{TRA}{TRA}{triple row activation}

\usepackage{xspace}
\usepackage[binary-units=true]{siunitx}
\usepackage{todonotes}
\usepackage{balance}
\usepackage{setspace}
\usepackage[export]{adjustbox}
\usepackage[us,12hr]{datetime}
\usepackage[en-GB, useregional=numeric]{datetime2}

\usepackage{fancyhdr}
\usepackage{cleveref}
\crefformat{section}{\S#2#1#3}
\crefformat{subsection}{\S#2#1#3}
\crefformat{subsubsection}{\S#2#1#3}

\newcommand{\versionnum}[0]{2.0}

\definecolor{MidnightBlue}{rgb}{0.1, 0.1, 0.44}
\definecolor{dollarbill}{rgb}{0.52, 0.73, 0.4}

\newif\ifcameraready
\camerareadytrue

\ifcameraready
    \newcommand{\gfcr}[1]{\textcolor{black}{#1}}

    \newcommand{\omcri}[1]{\textcolor{black}{#1}}
    \newcommand{\omcrii}[1]{\textcolor{black}{#1}}

\else 
    \newcommand{\gfcr}[1]{\textcolor{black}{#1}}

    \newcommand{\omcri}[1]{\textcolor{black}{#1}}
    \newcommand{\omcrii}[1]{\textcolor{red}{#1}}

    \let\marginpar\marginnote
\fi

\newcommand{\li}{(\textit{i})}
\newcommand{\lii}{(\textit{ii})}
\newcommand{\liii}{(\textit{iii})}
\newcommand{\liv}{(\textit{iv})}
\newcommand{\lv}{(\textit{v})}

\definecolor{blush}{rgb}{0.87, 0.36, 0.51}

\newcommand{\prop}{{DaPPA}\xspace}

\newcommand{\paratitle}[1]{\vspace{4pt}\noindent\textbf{#1.}}

\definecolor{airforceblue}{rgb}{0.36, 0.54, 0.66}
\definecolor{dodgerblue}{rgb}{0.12, 0.56, 1.0}
\definecolor{brandeisblue}{rgb}{0.0, 0.44, 1.0}
\definecolor{brickred}{rgb}{0.8, 0.25, 0.33}
\definecolor{eggplant}{rgb}{0.38, 0.25, 0.32}
\definecolor{byzantium}{rgb}{0.44, 0.16, 0.39}
\definecolor{ddgreen}{rgb}{0.00, 0.50, 0.00}

\definecolor{mygreen}{rgb}{0,0.6,0}
\definecolor{mygray}{rgb}{0.5,0.5,0.5}
\definecolor{mymauve}{rgb}{0.58,0,0.82}

\definecolor{bluehl}{rgb}{0.8,0.874,1}
\definecolor{pinkhl}{rgb}{0.992156863,0.847058824,1}
\definecolor{macaroniandcheese}{rgb}{1.0, 0.74, 0.53}
\definecolor{mossgreen}{rgb}{0.68, 0.87, 0.68}
\definecolor{greenhl}{rgb}{0.835,0.996,0.939}
\definecolor{yellowhl}{rgb}{0.996,0.957,0.8}
\definecolor{palecerulean}{rgb}{0.61, 0.77, 0.89}
\definecolor{gray(x11gray)}{rgb}{0.75, 0.75, 0.75}
\definecolor{amethyst}{rgb}{0.6, 0.4, 0.8}
\definecolor{ao}{rgb}{0.0, 0.5, 0.0}
\definecolor{burntorange}{rgb}{0.8, 0.33, 0.0}

\definecolor{cadmiumorange}{rgb}{0.93, 0.53, 0.18}

\definecolor{frenchlilac}{rgb}{0.53, 0.38, 0.56}
\definecolor{heliotrope}{rgb}{0.87, 0.45, 1.0}
\definecolor{peridot}{rgb}{0.9, 0.89, 0.0}
\definecolor{saffron}{rgb}{0.96, 0.77, 0.19}
\definecolor{tuscanred}{rgb}{0.51, 0.21, 0.21}
\definecolor{uscgold}{rgb}{1.0, 0.8, 0.0}
\definecolor{tangerineyellow}{rgb}{1.0, 0.8, 0.0}
\definecolor{rufous}{rgb}{0.66, 0.11, 0.03}
\definecolor{safetyorange}{rgb}{1.0, 0.4, 0.0}
\newcommand{\tempcommand}[1]{\renewcommand{\arraystretch}{#1}}

\newcommand\ignore[1]{ }
\newcommand{\revdel}[1]{}
\newcommand{\sgdel}[1]{}

\newif\ificcdsubmission
\iccdsubmissiontrue

\ificcdsubmission
    \newcommand{\gficcd}[1]{#1}
    \newcommand{\gficcdii}[1]{#1}
    \newcommand{\gficcdiii}[1]{#1}

\else
  \newcommand{\gficcd}[1]{{\color{blue}#1}}     
\newcommand{\gficcdii}[1]{{\color{red}#1}}
\newcommand{\gficcdiii}[1]{{\color{orange}#1}}
\fi

\newif\ifasplosrevisionsubmission
\asplosrevisionsubmissiontrue

\ifasplosrevisionsubmission
    
\else
    
\fi

\newif\ifasplosrevision
\asplosrevisionfalse
\ifasplosrevision 
    
\else

\fi

\newif\ifasplossubmission
\asplossubmissiontrue
\ifasplossubmission

    \newcommand{\agyasploscomment}[1]{}
\else

    \newcommand{\agyasploscomment}[1]{\textcolor{red}{\textbf{!!!~Giray:} #1}}
    
\fi

\newif\ifmicrosubmission
\microsubmissiontrue
\ifmicrosubmission

    \newcommand{\agymicrocomment}[1]{}
\else

    \newcommand{\agymicrocomment}[1]{\textcolor{red}{\textbf{!!!~Giray:} #1}} 
\fi

\newif\ifiscarevision
\iscarevisionfalse
\ifiscarevision 
    \newcommand{\revA}[1]{\textcolor{red}{#1}}
    \newcommand{\revB}[1]{\textcolor{mygreen}{#1}}
    \newcommand{\revC}[1]{\textcolor{safetyorange}{#1}}
    \newcommand{\revD}[1]{\textcolor{heliotrope}{#1}}
    \newcommand{\revE}[1]{\textcolor{rufous}{#1}}
    \newcommand{\revCommon}[1]{\textcolor{blue}{#1}}

    \newcommandx{\changeCM}[2][1=]{\todo[linecolor=blue,backgroundcolor=blue!25,bordercolor=blue,#1,size=\scriptsize]{\revCommon{\textbf{#2}}}}
    
    \newcommandx{\changeA}[2][1=]{\todo[linecolor=red,backgroundcolor=red!25,bordercolor=red,#1,size=\scriptsize]{\revA{\textbf{#2}}}}
    
    \newcommandx{\changeB}[2][1=]{\todo[linecolor=mygreen,backgroundcolor=mygreen!25,bordercolor=mygreen,#1,size=\scriptsize]{\revB{\textbf{#2}}}}
    
    \newcommandx{\changeC}[2][1=]{\todo[linecolor=safetyorange,backgroundcolor=safetyorange!25,bordercolor=safetyorange,#1,size=\scriptsize]{\revC{\textbf{#2}}}}
    
    \newcommandx{\changeD}[2][1=]{\todo[linecolor=heliotrope,backgroundcolor=heliotrope!25,bordercolor=heliotrope,#1,size=\scriptsize]{\revD{\textbf{#2}}}}
    
    \newcommandx{\changeE}[2][1=]{\todo[linecolor=rufous,backgroundcolor=rufous!25,bordercolor=rufous,#1,size=\scriptsize]{\revE{\textbf{#2}}}}
\else
    \newcommand{\revA}[1]{\textcolor{black}{#1}}
    \newcommand{\revB}[1]{\textcolor{black}{#1}}
    \newcommand{\revC}[1]{\textcolor{black}{#1}}
    \newcommand{\revD}[1]{\textcolor{black}{#1}}
    \newcommand{\revE}[1]{\textcolor{black}{#1}}
    \newcommand{\revCommon}[1]{\textcolor{black}{#1}}

    \newcommandx{\changeCM}[2][1=]{\todo[disable,#1]{#2}}
    \newcommandx{\changeA}[2][1=]{\todo[disable,#1]{#2}}
    \newcommandx{\changeB}[2][1=]{\todo[disable,#1]{#2}}
    \newcommandx{\changeC}[2][1=]{\todo[disable,#1]{#2}}
    \newcommandx{\changeD}[2][1=]{\todo[disable,#1]{#2}}
    \newcommandx{\changeE}[2][1=]{\todo[disable,#1]{#2}}
\fi

\newif\ifcut
\cutfalse

\ifcut
   \newcommand{\gfcut}[1]{} 
\else
    \newcommand{\gfcut}[1]{\textcolor{red}{\sout{#1}}}
\fi

\newif\ifiscasubmission
\iscasubmissiontrue

\ifiscasubmission
    
    \newcommand{\gfbisca}[1]{}

\else
    \newcommand{\gfbisca}[1]{\textcolor{blue}{\textit{GF: #1}}}

\fi

\newif\ifsubmission
\submissiontrue

\ifsubmission

    \newcommand{\jgl}[1]{}

    \newcommand{\gfb}[1]{}
    \newcommand{\mayank}[1]{}
    
    \newcommand{\agy}[1]{#1}
    \newcommand{\agycomment}[1]{}
\else
    \newcommand{\jgl}[1]{\textcolor{brickred}{\textit{JGL: #1}}}
    \newcommand{\gfb}[1]{\textcolor{blue}{\textit{GF: #1}}}
    \newcommand{\todo}[1]{\textcolor{red}{\textbf{TODO: #1}}}
    
    \newcommand{\mayank}[1]{\textcolor{green}{\textit{Mayank: #1}}}

    \newcommand{\agy}[1]{\textcolor{orange}{#1}}
    \newcommand{\agycomment}[1]{\agy{\textbf{[@gy:} #1\textbf{]}}}

\fi

\newcommand{\circled}[1]{\tikz[baseline=(char.base)]{\node[shape=circle,draw,inner sep=0pt,fill=black, text=white] (char) {#1};}}

\newcommand\pimdef{\cite{ghose.ibmjrd19, mutlu2020modern,deoliveira2021IEEE,pim-book,mutlu2019processing,mutlu2019enabling,mutlu2015research,mutlu2013memory,loh2013processing,Near-Data,stone1970logic,Miss_Mem_Wall_1996,Kautz1969,ahn2015scalable,mcciede2024,mutlu2025memory,PEI}\xspace}

\newcommand\pnm{\cite{farmahini2015nda,babarinsa2015jafar,devaux2019true,ghiasi2022genstore,gomez2021benchmarkingcut,gomez2022benchmarking,syncron,singh2020nero,skhynixpim,ke2021near,giannoula2022sparsep,shin2018mcdram,cho2020mcdram,denzler2021casper,asghari2016chameleon,IRAM_Micro_1997,C_RAM_1999,CASES_MVX,Xi_2015,sun2021abc,matam2019graphssd,gokhale1995processing,hall1999mapping,MEMSYS_MVX,lockerman2020livia,ahn2015scalable,nai2017graphpim,boroumand2018google,lazypim, top-pim, gao2016hrl, kim2018grim, drumond2017mondrian, RVU, NIM, PEI, gao2017tetris,Kim2016,gu2016leveraging, boroumand2019conda, hsieh2016transparent, cali2020genasm, NDC_ISPASS_2014,pattnaik2016scheduling,akin2015data,hsieh2016accelerating,lee2015bssync,boroumand2021mitigating,boroumand2021google,boroumand2022polynesia,boroumand2021polynesia,amiraliphd,besta2021sisa,fernandez2020natsa,singh2019napel,kwon202125,lee2021hardware,niu2022184qps,Sparse_MM_LiM,azarkhish2016logic,azarkhish2018neurostream,guo20143d,de2018design,akin2014hamlet,huang2020heterogeneous,dai2018graphh,liu2018processing,tsai:micro:2018:ams,gu2020ipim,DRAMA_CAL_2014,Asghari-Moghaddam_2016,huang2019active,kersey2017lightweight,li2019pims,kim2017grim,boroumand2017lazypim,zhuo2019graphq,zhang2018graphp,lim2017triple,smc_sim,HIVE,jang2019charon,IBM_ActiveCube,hadidi2017cairo,santos2018processing}\xspace}

\newcommand\drampum{\cite{angizi2019graphide,besta2021sisa,bostanci2022dr,deng2018dracc,ferreira2021pluto,ferreira2022pluto,gao2019computedram,li2017drisa,li2018scope,olgun2021quactrng,olgun2022pidram,seshadri.bookchapter17, seshadri2013rowclone,seshadri2015fast,seshadri2016buddy, seshadri2016processing, seshadri2017ambit,seshadri2018rowclone, seshadri2019dram, xin2020elp2im,missingnot,mimdramextended,oliveira2025proteus, oliveira2026pipedram, tokuda2026clutch, yuksel2025pudhammer, olgun2025dram, yuksel2024simultaneous}\xspace}

\newcommandx{\unsure}[2][1=]{\todo[linecolor=red,backgroundcolor=red!25,bordercolor=red,#1, size=\tiny]{#2}}
\newcommandx{\change}[2][1=]{\todo[linecolor=blue,backgroundcolor=blue!25,bordercolor=blue,#1,size=\tiny]{\textbf{#2}}}
\newcommandx{\feedback}[2][1=]{\todo[linecolor=yellow,backgroundcolor=yellow!25,bordercolor=yellow,#1]{#2}}
\newcommandx{\improvement}[2][1=]{\todo[linecolor=Plum,backgroundcolor=Plum!25,bordercolor=Plum,#1]{#2}}
\newcommandx{\thiswillnotshow}[2][1=]{\todo[disable,#1]{#2}}
\newcommandx{\completedRevision}[2][1=]{\todo[disable,backgroundcolor=red,#1]{#2}}
\newcommandx{\dataSource}[2][1=]{\todo[disable,backgroundcolor=red,#1]{#2}}
\newcommandx{\info}[2][1=]{\todo[linecolor=dollarbill,backgroundcolor=dollarbill!25,bordercolor=dollarbill,#1, size=\tiny]{#2}}

\newcommand{\boxbegin} {
	\begin{tcolorbox}[enhanced, frame hidden, colback=gray!50, breakable]
}

\newcommand{\boxend} {
	\end{tcolorbox}
}

\definecolor{lightblue}{rgb}{0.980, 0.956, 0.623}
\sethlcolor{lightblue}

\newcommand{\yboxbegin} {
	\begin{tcolorbox}[breakable, enhanced, frame hidden,
	enlarge top by=-0.25cm,
   enlarge bottom by=-0.1cm,
	colback=yellow!50]
}

\newcommand{\yboxend} {
	\end{tcolorbox}
}

\makeatletter
\patchcmd{\SOUL@ulunderline}{\dimen@}{\SOUL@dimen}{}{}
\patchcmd{\SOUL@ulunderline}{\dimen@}{\SOUL@dimen}{}{}
\patchcmd{\SOUL@ulunderline}{\dimen@}{\SOUL@dimen}{}{}
\newdimen\SOUL@dimen
\makeatother

\definecolor{backcolour}{rgb}{0.95,0.95,0.92}
\definecolor{mygreen}{rgb}{0,0.6,0}
\definecolor{mygray}{rgb}{0.5,0.5,0.5}
\definecolor{mymauve}{rgb}{0.58,0,0.82}
\definecolor{mblue}{rgb}{0.27,0.33,0.53}

\definecolor{codebg}{rgb}{1.0,1.0,1.0}

\definecolor{codebg}{rgb}{1.0,1.0,1.0}
\definecolor{standardkw}{rgb}{0.1,0.1,0.8}  
\definecolor{customkw}{rgb}{0.6,0.2,0.2}    
\definecolor{antiquewhite}{rgb}{0.98, 0.92, 0.84}
\definecolor{anti-flashwhite}{rgb}{0.95, 0.95, 0.96}
\definecolor{armygreen}{rgb}{0.29, 0.33, 0.13}

\lstdefinestyle{cppminted}{
    language=C++,
    basicstyle=\scriptsize\ttfamily,
    backgroundcolor=\color{anti-flashwhite},
    frame=lines,
    framexleftmargin=0mm,
    framexrightmargin=0mm,
    rulecolor=\color{black},
    framerule=0.03em,
    xleftmargin=1.5em,
    breaklines=true,
    postbreak=\mbox{\textcolor{gray}{$\hookrightarrow$\space}},  
    showstringspaces=false,
    numbers=left,
    numberstyle=\tiny\color{gray},
    stepnumber=1,
    commentstyle=\color{mygreen},    
    numbersep=5pt,
    keywordstyle=\color{standardkw}\bfseries,
    morekeywords={uint32_t},
    classoffset=1,
    morekeywords={Pipeline, stage, fetch, execute},
    keywordstyle=\color{customkw}\bfseries,
    classoffset=0
}

\lstdefinestyle{cppcustom}{
    language=C++,
    basicstyle=\scriptsize\ttfamily,
    backgroundcolor=\color{anti-flashwhite},
    frame=lines,
    framexleftmargin=0em,
    framexrightmargin=0em,
    rulecolor=\color{black},
    framerule=0.03em,
    xleftmargin=0em,
    breaklines=true,
    postbreak=\mbox{\textcolor{gray}{$\hookrightarrow$\space}},  
    showstringspaces=false,
    numbers=none,
    commentstyle=\color{mygreen},    
    keywordstyle=\color{standardkw}\bfseries,
    morekeywords={uint32_t},
    classoffset=1,
    morekeywords={Pipeline, stage, fetch, execute},
    keywordstyle=\color{customkw}\bfseries,
    classoffset=0
}

\ifcameraready
    \fancypagestyle{cameraready}{%
      \fancyhf{}%
      \fancyfoot[C]{\thepage}%
    }
\else
    \fancypagestyle{firstpage}{%
        \fancyhead{}%
        \fancyhead[C]{\textcolor{red}{CONFIDENTIAL DRAFT -- DO NOT DISTRIBUTE -- TO APPEAR IN ICCD'26} \\
                      \textcolor{MidnightBlue}{\emph{Version \versionnum~---~\today, \ampmtime}}}%
    }
\fi

\makeatletter
\def\bstctlcite{\@ifnextchar[{\@bstctlcite}{\@bstctlcite[@auxout]}}
\def\@bstctlcite[#1]#2{\@bsphack
  \@for\@citeb:=#2\do{%
    \edef\@citeb{\expandafter\@firstofone\@citeb}%
    \if@filesw\immediate\write\csname #1\endcsname{\string\citation{\@citeb}}\fi}%
  \@esphack}
\makeatother 

\begin{document}
\bstctlcite{IEEEexample:BSTcontrol}

\title{\scalebox{1.0}{\prop: \gficcd{A Unified} Data-Parallel Programming Framework}\\ \scalebox{1.0}{for Efficient Processing-in-Memory Execution}}

\author{
\vspace{5pt}
Geraldo F. Oliveira\textsuperscript{1}~\qquad 
Alain Kohli\textsuperscript{1}~\qquad 
David Novo\textsuperscript{2}~\qquad 
Ataberk Olgun\textsuperscript{1} \vspace{-5pt}\\ 
A. Giray Ya\u{g}l{\i}k\c{c}{\i}\textsuperscript{3}~\quad
Saugata Ghose\textsuperscript{4}~\qquad 
Juan Gómez-Luna\textsuperscript{5}~\qquad 
Onur Mutlu\textsuperscript{1, 6}\vspace{10pt}
\\
\textsuperscript{1}\emph{ETH Zürich} \qquad \qquad 
\textsuperscript{2}\emph{LIRMM, Univ. Montpellier, CNRS}\qquad \qquad 
\textsuperscript{3}\it{CISPA}\\
\textsuperscript{4}\emph{Univ. of Illinois Urbana-Champaign}\qquad \qquad
\textsuperscript{5}\emph{NVIDIA Research}\qquad \qquad
\textsuperscript{6}\emph{New York University}
\vspace{5pt}
}

\maketitle

\ifcameraready
  \thispagestyle{cameraready}
\else
  \thispagestyle{firstpage}
\fi

\begin{abstract}
\gficcd{\Gls{PIM} alleviates data-movement bottlenecks by placing computation near or inside memory.
Two general-purpose \gls{PIM} paradigms are now practical:
\gls{PnM} and \gls{PuM}.
Although both target regular, data-intensive, data-parallel workloads, they remain difficult to program because data organization is tightly coupled to computation and differs across substrates.
Thus, programmers must manually manage substrate-specific data partitioning, movement, layout, and tiling, preventing program \omcri{portability} across \gls{PIM} systems.}
\gficcd{We propose \prop{} (\underline{da}ta-\underline{p}arallel~\underline{p}rocessing-in-memory \underline{a}rchitecture),
a programming framework that exposes a single high-level data-parallel pattern interface and retargets it to multiple \gls{PIM} substrates.
The key idea of \prop{} is to decouple the expression of data parallelism from substrate-specific data organization: programmers express computation once using data-parallel patterns and a dataflow interface, while specialized \omcri{backends} automatically generate data orchestration for \gls{PnM} and multi-level tiling for \gls{PuM}.}
\gficcd{We evaluate \prop{} on representative substrates from both \gls{PIM} paradigms.
\gficcd{On the UPMEM \gls{PnM} system, \prop{} outperforms hand-tuned implementations by 1.6$\times$ on average with 95\% fewer lines-of-code, and outperforms the state-of-the-art SimplePIM framework by 1.4$\times$ on average with 62\% fewer lines-of-code.
On a SIMDRAM-like \gls{PuM} substrate, \prop{} outperforms \gficcdiii{the state-of-the-art MIMDRAM compiler} by 2.0$\times$ on average \gficcdiii{(up to 6.4$\times$)} across 12 workloads. \gficcdiii{\prop{} can deploy \emph{all} ten of our \gls{PnM} workloads, whereas CINM, the closest framework targeting both paradigms, can deploy \emph{only} three, on which \prop{} is 2.7--6.8$\times$ faster. \omcri{We open-source \prop{} at \url{https://github.com/CMU-SAFARI/DaPPA}.}}}}

\end{abstract}

\glsresetall

\glsresetall

\section{Introduction}
\label{sec:introduction}

\gficcd{Data movement between memory and processors dominates the execution time and energy consumption of modern data-intensive workloads in conventional processor-centric systems.
\Gls{PIM}~\omcri{\pimdef} alleviates this bottleneck by placing computation near or inside memory, and two general-purpose \gls{PIM} paradigms are now practical~\cite{mutlu2020modern}:
\li~\emph{\gls{PnM}}, which places programmable \gls{PIM} cores near DRAM banks and exposes \gficcdii{to the programmer} a C-like, thread-parallel programming model, as in the commercial UPMEM system~\omcri{\cite{gomez2022benchmarking, upmem2018, gomez2021benchmarkingcut, gomez2023evaluating}}; and
\lii~\emph{\gls{PuM}}, which performs bit-serial in-memory arithmetic operations across many DRAM columns using the analog operating principles of DRAM cells~\omcri{\drampum}, requiring operands in a transposed, bit-vertical data layout, as in SIMDRAM~\cite{hajinazarsimdram}.
Both \omcri{\gls{PnM} and \gls{PuM}} remain difficult to program because general-purpose \gls{PIM} tightly couples computation with \emph{data organization}~\omcri{\cite{yang2026dcc}}, i.e., how data is partitioned across compute units, moved through the memory hierarchy, and physically laid out within memory arrays.
This burden is also substrate-specific, since \gls{PnM} forces the programmer to \emph{explicitly} copy data \omcri{from/to} host memory, DRAM banks, and per-core scratchpads, whereas \gls{PuM} forces decisions about which operands to transpose~\omcri{\cite{oliveira2026pipedram}}, how to tile data across DRAM arrays, and which results to read back.
Consequently, a program written for one \gls{PIM} substrate \omcri{is \emph{not} portable} to another, even when both exploit the same \gficcdii{type of} data-level parallelism.}

\revdel{\gficcd{Existing \gls{PIM} programming frameworks~\cite{mimdramextended, chen2023simplepim, khan2022cinm} do \emph{not} comprehensively address these challenges, instead leaving programmers with fragmented, substrate-specific abstractions.
For \gls{PnM}, low-level vendor software development kits (SDKs) expose the full hardware interface and require \emph{manual} computation partitioning and data movement orchestration, while higher-level frameworks, such as SimplePIM~\cite{chen2023simplepim}, offer only a small set of computing primitives (i.e., \texttt{map}, \texttt{reduce}, and \texttt{zip}) and still require explicit communication management from the programmer.
For \gls{PuM}, frameworks such as MIMDRAM~\cite{mimdramextended} rely on compiler loop auto-vectorization, which \emph{cannot} effectively handle nested loops and thus underutilizes the available in-memory resources.
Cinnamon~\cite{khan2022cinm} targets both paradigms through an MLIR~\cite{mlir}-level interface built around a fixed catalog of predefined operators, such as matrix multiplication, convolution, and reduction, but requires the user to either write \emph{non-trivial} MLIR dialects or lower application code into \emph{limited, predefined} linear-algebra-centric ones, raising yet another programmability barrier.}}

\gficcd{Our \emph{goal} in this work is to ease and unify programmability across general-purpose \gls{PIM} substrates, enabling a programmer to express a computation once and efficiently target different \gls{PIM} substrates \emph{without} explicitly managing hardware-specific data organization.
To this end, we propose \prop{}
(\underline{da}ta-\underline{p}arallel~\underline{p}rocessing-in-memory \underline{a}rchitecture), a programming framework that exposes a single high-level data-parallel pattern interface and \omcri{targets} it to different \gls{PIM} substrates.
The key idea of \prop{} is to decouple the computation's data parallelism, which the programmer captures once using high-level C++-based data-parallel patterns, from the substrate-specific data organization, which specialized \omcri{backends} generate \omcri{\emph{automatically}}.
This decoupling enables the same \omcri{frontend} program to drive substrate-specific code generation, eliminating the need for the programmer to manually write the data orchestration, tiling, or transposition logic required by each substrate. \gficcdiii{\prop is a \emph{pattern-based} programming framework, \emph{not} an automatic compiler for arbitrary code: the programmer expresses the computation as a pipeline of data-parallel patterns whose bodies hold arbitrary C++ code, and \prop generates the substrate-specific data organization that this pipeline requires.}}

\gficcd{\prop{} consists of three components.
First, \prop's \emph{data-parallel pattern APIs} are a collection of primitives (\texttt{map}, \texttt{filter}, \texttt{reduce}, \texttt{scan}, \texttt{window}, and \texttt{group}) for expressing how data is transformed within an application.
Second, \prop's \emph{dataflow programming interface} specifies how data moves across data-parallel patterns by organizing them into an ordered \omcri{set of} pipeline of stages.
Third, \prop's \emph{dynamic template-based compilation} automatically generates the data organization required by each substrate, including non-unified data orchestration for \gls{PnM} and multi-level tiling and transposition for \gls{PuM}, and transparently emits and compiles substrate-specific code.}

\gficcd{We evaluate \prop{} on representative substrates from both \gls{PIM} paradigms.
On the UPMEM \gls{PnM} system~\omcri{\cite{gomez2022benchmarking, upmem2018, gomez2021benchmarkingcut, gomez2023evaluating}}, across ten workloads that exercise all six data-parallel patterns, \prop{} outperforms the hand-tuned implementations from the PrIM benchmark suite~\omcri{\cite{gomezluna2021repo,gomez2022benchmarking,gomez2021benchmarkingcut}} by 1.6$\times$ on average (up to 6.0$\times$) with 95\% fewer lines-of-code, and is never slower on any workload.
Compared to SimplePIM~\omcri{\cite{chen2023simplepim, simplepimrepo}}, a state-of-the-art \gls{PnM} programming framework, \prop{} is 1.4$\times$ faster on average (up to 2.2$\times$) with 62\% fewer lines-of-code. \gficcdiii{Compared to CINM~\cite{khan2022cinm}, the closest framework that also targets both \gls{PIM} paradigms, \prop{} expresses \emph{all} ten of our \gls{PnM} workloads, whereas CINM expresses \emph{only} three, and \prop{} is 2.7--6.8$\times$ faster on those three.}
On a SIMDRAM-like \gls{PuM} substrate~\cite{hajinazarsimdram}, \omcri{\prop{}} outperforms \gficcdiii{the state-of-the-art MIMDRAM compiler~\cite{mimdramextended}} by 2.0$\times$ on average (up to 6.4$\times$) across 12 data-parallel kernels\revdel{, since \prop's back-end keeps the substrate's wide subarray datapath busy where inner-loop vectorization leaves most columns idle}.}

\omcri{This work makes the following major contributions:}
\begin{itemize}[noitemsep,topsep=0pt,parsep=0pt,partopsep=0pt,labelindent=0pt,itemindent=0pt,leftmargin=*]
    \item \omcri{We identify the central programmability challenge across general-purpose \gls{PIM} substrates: data organization is tightly coupled to computation, and the required \omcri{data} organization differs substantially across \omcri{\gls{PnM} and \gls{PuM}} substrates.}

    \item \omcri{We propose \prop{}, a programming framework that exposes a single source-level data-parallel pattern interface and targets it to both \gls{PnM} and \gls{PuM} substrates.}

    \item \omcri{We design specialized \gls{PIM} \omcri{backends} that \emph{automatically} generate the data organization required by each \gls{PIM} substrate, including non-unified data orchestration for \gls{PnM} and multi-level tiling and data transposition for \gls{PuM}.}
    
    \item \omcri{We evaluate \prop{} on representative \gls{PnM} and \gls{PuM} substrates, showing that it eases and unifies programmability while matching or exceeding the performance of hand-tuned code and prior \gls{PIM} programming frameworks.}

    \item \omcri{We open-source \prop{} at \url{https://github.com/CMU-SAFARI/DaPPA}.}
\end{itemize}

\section{\omcri{Background, Motivation, and Our Goal}}
\label{sec:background}

\subsection{\gficcd{General-Purpose PIM Systems}}

\begin{figure}[ht]
    \centering
    \includegraphics[width=\linewidth]{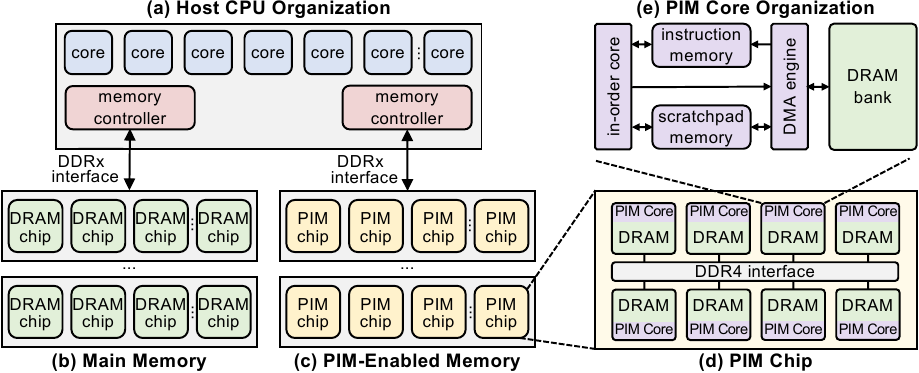}
    \caption{A general-purpose \gls{PnM} system organization.}
    \label{fig:architecture}
\end{figure}

\paratitle{\gficcd{Processing-Near-Memory (\gls{PnM})}}
\gficcd{In \gls{PnM}, small general-purpose in-order cores are placed next to DRAM banks, so computation runs on conventional cores with high-bandwidth, low-latency access to memory~\omcri{\pnm}.
UPMEM~\omcri{\cite{upmem2018, gomez2022benchmarking,gomez2021benchmarkingcut,chen2023simplepim,giannoula2024pygim,giannoula2022sparsep,gomez2023evaluating,syncron,item2023transpimlib,lee2024pimmmu,hyun2024pathfinding}} is the first commercially available general-purpose \gls{PnM} architecture; Fig.~\ref{fig:architecture} illustrates its organization.
A \gls{PnM} system consists of a host CPU (Fig.~\ref{fig:architecture}a) with conventional DRAM main memory (Fig.~\ref{fig:architecture}b) and \gls{PIM}-enabled DDRx DIMMs (Fig.~\ref{fig:architecture}c), each populated with multiple \gls{PIM} chips (Fig.~\ref{fig:architecture}d).
Each \gls{PIM} core (Fig.~\ref{fig:architecture}e) has exclusive access to
\li~a DRAM bank, around \SI{64}{\mega\byte} in size;
\lii~an instruction memory, around \SI{24}{\kilo\byte} in size; and
\liii~a scratchpad memory, around \SI{64}{\kilo\byte} in size.
Crucially for programming, a \gls{PnM} system exposes a \emph{non-unified} memory space: the host main memory, each core's DRAM bank, and each core's scratchpad are separate address spaces, and cores attached to different banks can neither share memory nor communicate directly.}

\paratitle{\gficcd{Processing-Using-Memory (\gls{PuM})}}
\gficcd{In DRAM-based \gls{PuM}~\omcri{\drampum}, computation is performed using the analog operating principles of DRAM cells, allowing memory arrays to both \emph{store} and \emph{compute} on data.
General-purpose \gls{PuM} architectures, such as SIMDRAM~\cite{hajinazarsimdram} and MIMDRAM~\cite{mimdramextended}, build arbitrary in-DRAM arithmetic operations from a set of bulk-bitwise primitives (e.g., majority, NOT, and row copy~\omcri{\cite{seshadri2017ambit,seshadri2013rowclone, seshadri2015fast}}) and execute them \emph{bit-serially} across the columns of a DRAM subarray.
Since one in-DRAM operation acts on \emph{every} column of the subarray at \emph{once} (e.g., over 65,536 DRAM columns in a DDR4 DRAM module~\cite{micron2014ddr4}), a single subarray provides a very-wide \gls{SIMD} datapath for \gls{PuM} operations.
To perform in-DRAM arithmetic at low cost, operands must be stored in a \emph{vertical} data layout, where the multiple bits of a data word are distributed across a single DRAM column, allowing bit position $i$ of all elements to be computed together\gficcdii{. The} host sequences the bit-serial operations, while a data transposition unit converts data between the horizontal layout and the vertical layout~\cite{hajinazarsimdram,mimdramextended,oliveira2025proteus}.
Because computation happens in place, a \gls{PuM} substrate exposes a \emph{unified} memory space in which data is computed where it resides.}

\revdel{\begin{figure}[ht]
    \centering
    \includegraphics[width=0.95\linewidth]{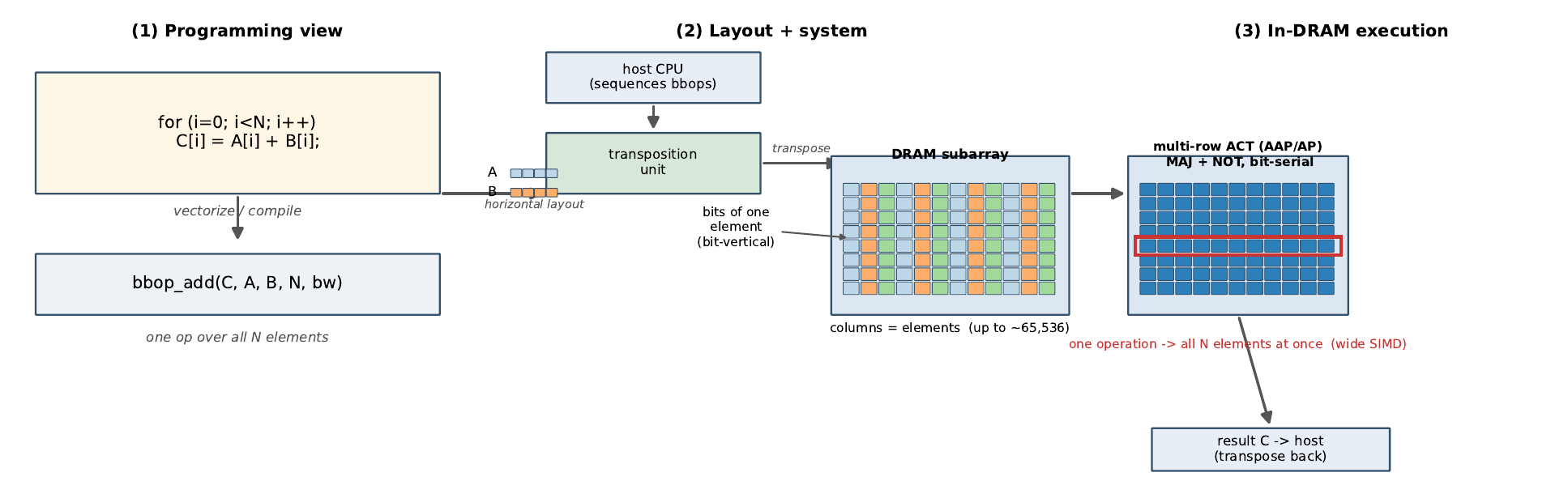}
    \caption{A general-purpose \gls{PuM} system, using SIMDRAM as an example, executing a vector addition: a vectorized bulk-bitwise operation runs bit-serially across all subarray columns at once.}
    \label{fig:pum}
\end{figure}}

\subsection{\gficcd{The Data Organization \omcri{Challenge} of General-Purpose \gls{PIM}}}

\gficcd{Programming general-purpose \gls{PIM} systems requires the programmer/compiler to explicitly manage \emph{data organization}, which includes how data is
\li~partitioned across compute units,
\lii~moved through the memory hierarchy, and
\liii~physically laid out within memory arrays.
Both \gls{PnM} and \gls{PuM} impose these three \omcri{challenges} in different \omcri{ways}.}

\noindent \underline{Partitioning:} \gficcd{\omcri{In} \gls{PnM}, the programmer partitions both the computation and its input data across many \gls{PIM} cores and across the \gls{PIM} threads within each core, manually exposing thread-level parallelism.
\omcri{In} \gls{PuM}, the programmer maps the computation onto subarray columns and, when the data exceeds a subarray's width or a single operation's bit-width, applies multi-level tiling across columns and subarrays.}

\noindent \underline{Data Movement:} \gficcd{\omcri{In} \gls{PnM}, the non-unified memory space forces the programmer to explicitly copy data from host memory into DRAM banks and then stage operands between each bank and its local scratchpad memory, under strict address-alignment and transfer-size constraints.
\omcri{If} \gls{PIM} cores \emph{cannot} communicate directly \omcri{(in some existing systems~\cite{upmem, gomez2022benchmarking, gomez2021benchmarkingcut})}, any inter-core interaction, such as merging partial results, must be routed through the host CPU, and within a \gficcdii{PIM} core, the programmer must manage synchronization (e.g., mutexes, barriers).
\omcri{In} \gls{PuM}, computation is \emph{in place}, so data movement instead takes the form of deciding which operands to convert into the vertical data layout, moving operands across subarrays when an operation spans more than one subarray, and reading back and transposing the results consumed by the host CPU.}

\noindent \underline{Data Layout:} \gficcd{\omcri{In} \gls{PnM}, \gficcdii{\emph{all} input/output} arrays must satisfy the alignment and granularity constraints of the \gficcdii{DRAM} bank and scratchpad memories.
\omcri{In} \gls{PuM}, every \gls{PuM} operand must be vertically transposed and tiled to fit the subarray dimensions, and results must be transposed back to the conventional horizontal data layout when accessed by the host CPU.}

\paratitle{Problem and Goal}
\gficcd{Across both \omcri{\gls{PIM}} paradigms, data organization is inseparable from computation and differs substantially across substrates, so programming a general-purpose \gls{PIM} system requires detailed hardware knowledge, and a program written for one substrate \omcri{is not portable} to another.
Existing \gls{PIM} programming systems do \emph{not} eliminate this burden.
SimplePIM~\cite{chen2023simplepim}\omcri{, designed for \gls{PnM},} raises the \gficcdii{programming} abstraction level \omcri{in} \gls{PnM}\gficcdii{,} yet\gficcdii{, it} still requires explicit communication management and targets a single substrate.
MIMDRAM~\cite{mimdramextended}\omcri{, a \gls{PuM} substrate,} relies on loop auto-vectorization that extracts parallelism only from the innermost loop, leaving most \gls{PuM} columns idle for nested-loop kernels.
Cinnamon~\cite{khan2022cinm}\omcri{, spanning  both \gls{PnM} and \gls{PuM} paradigms,} exposes an MLIR~\cite{mlir}-level interface around a fixed, linear-algebra-centric operator set, requiring the programmer to learn \emph{non-trivial} compiler dialects.
\omcri{Our} \textbf{goal} is to ease and unify programmability across general-purpose \omcri{\gls{PnM} and \gls{PuM}} substrates, allowing a programmer to express a computation \omcri{only} \gficcdii{using a familiar programming interface} \emph{without} explicitly managing hardware resources \omcri{in either \gls{PnM} and \gls{PuM}}.}

\section{\prop Overview}
\label{sec:overview}

\revdel{To ease the programmability of PIM architectures, we propose \prop (\underline{da}ta-\underline{p}arallel \underline{p}rocessing-in-memory \underline{a}rchitecture), a programming framework that can, for a given application, \emph{automatically} distribute input and gather output data, handle memory management, and parallelize work across the DPUs. 
The \emph{key idea} behind \prop is to remove the responsibility of managing hardware resources from the programmer by providing an intuitive data-parallel pattern-based programming interface~\cite{cole1989algorithmic} that abstracts the hardware components of the UPMEM system. 
Using this key idea, \prop transforms a data-parallel pattern-based application code into the appropriate UPMEM-target code, including the required APIs for data management and code partition, which can then be compiled into a UPMEM-based binary \emph{transparently} from the programmer. 
While generating UPMEM-target code, \prop implements several code optimizations to improve performance.}

\begin{figure}[ht]
    \centering
    \includegraphics[width=\linewidth]{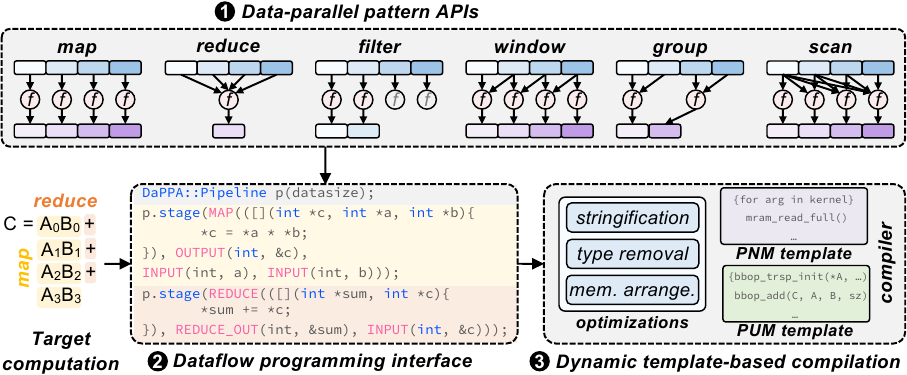}
    \caption{Overview of the \prop programming framework.}
    \label{fig:framework_overview}
\end{figure}

Fig.~\ref{fig:framework_overview} shows an overview of \omcri{the} \prop programming framework. \prop takes as input C/C++ code, which describes the target computation using a collection of data-parallel patterns and \prop's programming interface, and generates as output the requested computation. \prop consists of three main components:
\li~\prop's data-parallel pattern APIs,
\lii~\prop's dataflow programming interface, and
\liii~\prop's dynamic template-based compilation.

\paratitle{Data-Parallel Pattern APIs} \prop's data-parallel pattern APIs (\circled{1} in Fig.~\ref{fig:framework_overview}) are a collection of pre-defined function skeletons that implement high-level data-parallel pattern primitives. 
\omcri{Using these}, programmers can focus on application logic rather than low-level concerns such as \gls{PIM} thread management, synchronization, or inter-\gls{PIM} core communication. 
Each primitive allows the user to express data transformations. 
\revdel{\prop supports five primary data-parallel pattern primitives, including:
\li~\texttt{map}, which applies a function $f$ to each individual input element $i$, producing unique output elements $y_i = f(x_i)$;
\lii~\texttt{reduce}, which reduces input vectors to a scalar;
\liii~\texttt{filter}, which selects input elements based on a predicate;
\liv~\texttt{window}, which \emph{maps} and output element as the \emph{reduction} of $W$ \emph{overlapping} input elements; 
\lv~\texttt{group}, which \emph{maps} and output element as the \emph{reduction} of $G$ \emph{non-overlapping} input elements.  The user can combine these five data-parallel primitives to describe complex data transformations in an application.} 

\paratitle{Dataflow Programming Interface} \prop exposes a dataflow programming interface to the user (\circled{2} in Fig.~\ref{fig:framework_overview}). In this programming interface, the main component is the \texttt{Pipeline} class, which represents a sequence of data-parallel patterns that will be executed on the \gls{PIM} hardware. 
A given \texttt{Pipeline} has one or more \texttt{stage}s. 
Each \texttt{stage} \omcri{uses} a given data-parallel pattern primitive to transform input operands following a user-defined computation sequence. \texttt{Stage}s are executed in order, in a \omcri{pipelined} fashion.

\paratitle{Dynamic Template-Based Compilation} \gficcd{\prop uses dynamic template-based compilation (\circled{3} in Fig.~\ref{fig:framework_overview}) to generate substrate-specific \gls{PIM} code from a single \omcri{frontend} program.
\prop pairs each target substrate \omcri{(\gls{PnM} or \gls{PuM})} with a back-end template that captures how that substrate organizes data, and specializes the template to the user's pipeline in two main steps.
In the first step, \prop creates an initial \gls{PIM} code from the \omcri{backend} template of the target substrate.
In the second step, \prop applies a series of transformations to
\li~extract from the user program the information needed to specialize the template;
\lii~generate the data organization that the target substrate requires, i.e., orchestration and offset computation across DRAM banks and scratchpad memories for \gls{PnM}, and multi-level tiling and transposition into the bit-vertical layout for \gls{PuM}; and
\liii~divide computation between the host and the \gls{PIM} hardware.
\prop then transparently emits and compiles the resulting code into a substrate-specific \gls{PIM} binary.}

\revdel{\paratitle{Putting All Together} Using \prop's data-parallel pattern APIs, dataflow programming interface, and dynamic template-based compilation, the user can quickly implement and deploy applications to the UPMEM system without any knowledge of the underlying architecture. Figure~\ref{fig:framework_overview} showcases an example of implementing a simple vector dot-product application using \prop. In this example, the user defines a \texttt{Pipeline} with two \texttt{stage}s: a \texttt{map} stage and a \texttt{reduce} stage. \prop generates the appropriate binary for the UPMEM system, executes the target computation in the DPUs, and copies the final output from the DPUs to the CPU.}

\section{\prop Implementation}
\label{sec:implementation}

\omcri{We describe our implementation of \prop's three components. We open-source \prop at \url{https://github.com/CMU-SAFARI/DaPPA}.} 

\subsection{Data-Parallel Pattern APIs}

\prop exposes \emph{data-parallel pattern APIs} that leverage skeleton-based programming and data-parallel patterns to allow users to define their algorithms using high-level abstractions~\cite{cole1989algorithmic}. 
\gficcd{Currently, \prop supports six primary data-parallel patterns, i.e., \texttt{map}, \texttt{reduce}, \texttt{scan}, \texttt{filter}, \texttt{window}, and \texttt{group}\omcri{.} Table~\ref{tab:patterns} summarizes their syntax and semantics. Each pattern takes one or more one-dimensional (1D) vectors and produces a single 1D vector or a scalar, while non-vector arguments such as scalar parameters are broadcast across all participating PIM cores.\footnote{We use the term \emph{PIM core} independently \omcri{of the \gls{PIM} substrate} to denote the unit of data-parallel execution\omcri{. A \gls{PIM} core defines} a near-bank core and its threads \omcri{in} \gls{PnM}, and a DRAM column that acts as a SIMD lane \omcri{in} \gls{PuM}.}}

\begin{table}[ht]
\centering
\caption{\prop's six data-parallel pattern APIs. \(N\) is the input vector length, and \(f\) is a user-supplied pure function.}
\label{tab:patterns}
\footnotesize
\setlength{\tabcolsep}{3.5pt}
\begin{tabular}{@{}l l p{0.50\linewidth}@{}}
\toprule
\textbf{Pattern} & \textbf{Syntax} & \textbf{Semantics} \\
\midrule
\texttt{map}    & \texttt{map(f,y,x)}      & \(y_i = f(x_i)\); each output is computed independently. \\
\texttt{reduce} & \texttt{reduce(f,r,x)}   & Scalar \(r = f(x_1,\dots,x_N)\) with \(f\) associative. \\
\texttt{scan}   & \texttt{scan(f,y,x)}     & Inclusive prefix \(y_i = f(x_1,\dots,x_i)\) with \(f\) associative. \\
\texttt{filter} & \texttt{filter(f,y,x)}   & \(y\) keeps each \(x_i\) with \(f(x_i)=\texttt{true}\), preserving order. \\
\texttt{window} & \texttt{window(f,y,x,W)} & \(y_i = f(x_i,\dots,x_{i+W-1})\), giving \(M = N-W+1\) overlapping windows of width \(W\). \\
\texttt{group}  & \texttt{group(f,y,x,G)}  & \(y_i = f\) over the \(i\)-th of \(N/G\) disjoint size-\(G\) chunks. \\
\bottomrule
\end{tabular}
\end{table}

\gficcd{All pattern functions are \emph{pure}, i.e., deterministic and free of side effects, so each element or group is processed independently without synchronization or data sharing. \texttt{reduce} and \texttt{scan} further require \(f\) to be associative, which lets each PIM core compute a partial result that the host CPU folds into the final output, and \texttt{window} requires the user to pad the input so \omcri{that} the trailing positions form a complete window.}
\gficcd{\prop also allows the user to combine some of the six primary data-parallel patterns to implement more complex execution patterns.} In our current implementation, \prop enables the combination of \texttt{window}, \texttt{filter}, and \texttt{group} data-parallel patterns into four new implementations: \texttt{window+group}, \texttt{window+filter}, \texttt{group+filter}, and \texttt{window+group+filter}. 

\subsection{Dataflow Programming Interface}

\prop allows the programmer to compose the data-parallel pattern APIs into a complete application \omcri{via} a \emph{dataflow programming interface}. The \emph{key idea} is to let the programmer \emph{implicitly} describe the data movement between data-dependent sub-tasks. A \prop program is a \texttt{Pipeline}, an ordered collection of \texttt{stage}s in which each \texttt{stage} is a single data-parallel pattern, and data flows sequentially from one \texttt{stage} to the next. The output of a \texttt{stage} becomes the input of any later \texttt{stage}, and while defining a \texttt{stage} the user writes \emph{no} PIM-specific command for data orchestration or parallelism distribution, since \prop abstracts the underlying hardware. \gficcd{Because the interface is substrate-independent, the \emph{same} \texttt{Pipeline} program is retargeted to different \gls{PIM} substrates by \prop's \omcri{backend}s.}
The user builds an application in five steps, which we illustrate with the vector dot-product \omcri{example} in Listing~\ref{listing:vecdot}.
First, the user allocates the input arrays, supplied as either \texttt{std::vector}s or raw pointers (lines~\ref{ln:alloc}--\ref{ln:fill}).
Second, the user creates a \texttt{Pipeline} and specifies the \texttt{length} of its vectors (line~\ref{ln:pipe}); this length must be the same across all \texttt{stage}s, and computations needing different lengths are expressed by combining several \texttt{Pipeline}s.
Third, the user adds the \texttt{stage}s (lines~\ref{ln:map}--\ref{ln:stageend}). 
Each \texttt{stage} takes a macro naming its data-parallel pattern (e.g., \texttt{MAP}, \texttt{REDUCE}), a \emph{pure} per-thread kernel \texttt{func}, and a list of typed arguments that flag each one as an input, output, in-out, scalar parameter, scalar reduction output (\texttt{REDUCE\_OUT}), or cross-core combination function, together with the \texttt{overlap} or \texttt{groupSize} count used by \texttt{window} and \texttt{group}. \omcri{In the example}, a \texttt{map} multiplies $a$ and $b$ into the intermediate vector $c$ (line~\ref{ln:map}) and a \texttt{reduce} sums $c$ into the scalar $sum$ (line~\ref{ln:reduce}).

Fourth, the user marks with \texttt{fetch} the outputs to copy back to the CPU (line~\ref{ln:fetch}). An array used \emph{only} as intermediate data between \texttt{stage}s, such as $c$, is never fetched and therefore needs no host CPU allocation, which saves host memory.
Fifth, the user calls \texttt{execute} (line~\ref{ln:exec}), at which point \prop allocates the PIM cores, runs every \texttt{stage}, and writes the fetched results (i.e., $sum$) back to the CPU. \gficcd{When a \texttt{Pipeline} contains a \texttt{filter}, the user afterwards calls \texttt{getLength} to obtain the output length, which cannot be known \emph{a priori}; for all other patterns the length is computed in advance.}

\begin{figure}[ht]
\setcaptiontype{lstlisting}
\linespread{1.0}\selectfont
\centering
\begin{lstlisting}[style=cppminted,escapeinside={(*@}{@*)}]
uint32_t dataLength = 1048576; // input size
std::vector<uint32_t> a(dataLength); // input array (*@\label{ln:alloc}@*)
std::vector<uint32_t> b(dataLength); // input array
uint32_t c; // intermediate data between stages
uint32_t sum = 0; // output scalar
// Data initialization routine for input arrays a and b (*@\label{ln:fill}@*)

DaPPA::Pipeline p(dataLength);(*@\label{ln:pipe}@*)
// element-wise multiplication of input arrays a and b
p.stage(MAP(([](uint32_t *c, uint32_t *a, uint32_t *b){(*@\label{ln:map}@*)
    *c = *a * *b;
}),
    OUTPUT(uint32_t, &c),
    INPUT(uint32_t, a),
    INPUT(uint32_t, b)));

// reduce temporary array produced by prior stage into a scalar 
p.stage(REDUCE(([](uint32_t *sum, uint32_t *c){(*@\label{ln:reduce}@*)
    *sum += *c;
}),
    REDUCE_OUT(uint32_t, &sum),
    INPUT(uint32_t, &c)));(*@\label{ln:stageend}@*)

p.fetch(&sum);(*@\label{ln:fetch}@*)
p.execute();(*@\label{ln:exec}@*)
\end{lstlisting}
\caption{\omcri{An} implementation of vector dot-product using \omcri{\prop's} dataflow programming interface.}
\label{listing:vecdot} 
\end{figure}

\subsection{Dynamic Template-Based Compilation}

\gficcd{After the user implements the target computation using the data-parallel pattern APIs and the dataflow programming interface, \prop generates the appropriate \gls{PIM} binary using a \emph{dynamic template-based compilation} \omcri{process} that \omcri{has} two phases.
\omcri{The} first phase, a \omcri{\emph{substrate-independent pipeline analysis}}, common to all substrates, extracts the computation from the user's \texttt{Pipeline} into a substrate-independent intermediate form.
\omcri{The} second phase, a \emph{substrate-specific \omcri{backend}} specializes this intermediate form to the target substrate, where each \omcri{backend implements} a code \emph{template} that captures how the substrate organizes data with a sequence of \emph{transformations} that populate the template.
\prop currently provides two \omcri{backends}, one for \gls{PnM} (targeting the UPMEM system) and one for \gls{PuM} (targeting a SIMDRAM substrate), and repeats the whole process dynamically per \texttt{Pipeline}.
We use the Inja~\cite{inja} template engine to populate the code templates.}

\subsubsection{\omcri{Substrate-Independent Pipeline Analysis}}

\gficcd{The \omcri{substrate-independent pipeline analysis} extracts, from each \texttt{stage} of the \texttt{Pipeline}, the kernel function \texttt{func} and its input/output arguments, adding the kernels to a \emph{global stage list} and the arguments, with their input/output flags, to a \emph{global argument list}. It also strips most of the argument type information, keeping only the type size and its string representation, which reduces the templating effort and the compilation time. Both \omcri{backends} consume these two lists, so this step runs once per \texttt{Pipeline}, independently of the target substrate.}

\subsubsection{\gficcd{\gls{PnM} \omcri{Backend}}}
\label{sec:impl:pum}
\gficcd{The \gls{PnM} \omcri{backend} generates code for the UPMEM system~\omcri{\cite{gomez2022benchmarking, upmem2018, gomez2021benchmarkingcut, gomez2023evaluating}} using \emph{template-based coding}, starting from an initial skeleton of a \gls{PIM} application together with the global stage and argument lists, and applying three code transformations \omcrii{(T1, T2, T3)} that produce a \gls{PIM} binary.}

\paratitle{\omcrii{\underline{T1}: Memory Arrangement}} \gficcd{\omcrii{This} transformation computes the parameters that require a global view of the \texttt{Pipeline}, i.e., how data is arranged across the DRAM banks and scratchpad memories.
\prop splits the input data \emph{evenly} across all \gls{PIM} cores and, within a core, evenly across the \gls{PIM} threads, respecting the 8-byte alignment requirement, and places each argument at a computed offset from the DRAM bank base address.
The data for all \texttt{stage}s is copied \omcri{into the DRAM bank(s)} before execution starts, and when it exceeds the combined DRAM bank capacity, \prop processes it in multiple execution rounds; intermediate results that propagate through the \texttt{Pipeline} receive their own DRAM bank space.
At \gficcdii{runtime}, each \texttt{stage} stages data from the DRAM bank into the scratchpad memory in blocks sized to fit, casting each pointer to its argument type before invoking the kernel and writing the results back.}

%

\paratitle{\omcrii{\underline{T2}: Memory Alignment-Aware Work Distribution}} \gficcd{\omcrii{This} transformation handles data that does \emph{not} satisfy the 8-byte alignment requirement of the \gls{PIM} system. \prop generates host code that processes these leftover elements on a separate CPU thread while the \gls{PIM} cores process the aligned portion, overlapping the \omcri{execution of the two portions}.}

\paratitle{\omcrii{\underline{T3}: Host-Side Post-Processing}} \gficcd{\omcri{\omcrii{This} transformation generates the host code required to assemble the final outputs of \texttt{stage}s containing a \texttt{filter}, a \texttt{reduce}, or a \texttt{scan} data-parallel pattern.}
\omcri{For a \texttt{filter}, each \gls{PIM} thread writes the elements that satisfy the predicate of the filter \emph{contiguously} into its designated portion of the \gls{PIM} core's output buffer and records the number of elements written.
Since \prop transfers the entire output buffer from each \gls{PIM} core to the host, the transferred buffer can contain unused space after the valid output produced by each \gls{PIM} thread.
On the host, \prop uses the per-PIM-thread output counts to copy the valid elements from each PIM thread's portion into a contiguous output vector.}
\omcri{For a \texttt{reduce}, \prop combines the partial results produced by the individual \gls{PIM} cores on the host.
For a \texttt{scan}, \prop adjusts and concatenates the local scans using the offset computed for each \gls{PIM} core.}}
\gficcd{\omcri{After generating the required host-side code,} \prop inserts all parameters and kernels into the \omcri{code template} and compiles the resulting \gls{PIM} code \gficcdii{using the UPMEM compiler}.}

\subsubsection{\gficcd{\gls{PuM} \omcri{Backend}}}

\gficcd{The \gls{PuM} \omcri{backend generates code for \omcrii{the} SIMDRAM substrate~\omcrii{\cite{hajinazarsimdram}} using} a \gls{PuM} template together with the global stage and argument lists \omcri{produced by the substrate-independent pipeline analysis}.
\omcri{The backend applies three code transformations \omcrii{(T1, T2, T3) below} that translate the \texttt{Pipeline} into a schedule of bulk-bitwise operations executed bit-serially across the columns of a subarray.
The MIMDRAM compiler~\cite{mimdramextended,mimdramgit} then compiles this schedule into the final \gls{PIM} binary.}}

\paratitle{\omcrii{\underline{T1}: Operand Transposition}} \gficcd{\omcrii{This} transformation manages \emph{data layout} by emitting SIMDRAM's specialized data transposition \emph{bbop} instructions, which register each vector operand with the hardware data transposition unit that converts it from the conventional horizontal layout into the bit-vertical layout that in-DRAM computation requires. A scalar parameter is first replicated into an array, with one copy per column, and transposed in the same way, so that every column holds the parameter's bits in the bit-vertical layout.}

\paratitle{\omcrii{\underline{\omcrii{T2}}: Subarray Tiling \& Packing}} \gficcd{\omcrii{This} transformation manages \emph{tiling} \omcri{and \gls{SIMD} lane packing}. A single in-DRAM operation computes on all columns of one subarray at once, up to the subarray width $W$ of tens of thousands of columns. 
Therefore, \prop maps each operation onto a set of subarray-wide tiles by assigning the elements that are computed in parallel to columns, padding their count up to a multiple of $W$, and issuing one bulk-bitwise operation per tile, so an operation over $n$ parallel elements becomes $\lceil n/W\rceil$ subarray-wide operations, with a further level of tiling across subarrays when $n$ exceeds the columns of a single subarray.}
\omcri{A central decision in the second transformation is how to use the subarray columns efficiently.
Each \texttt{stage} produces $N_{out}$ independent outputs, and computing each output requires combining $G$ inputs.
\prop can map either the $G$ inputs of one output or multiple independent outputs across the columns, which we refer to as \emph{reduction-in-lane} and \emph{output-stationary} mappings, respectively.
The \emph{reduction-in-lane mapping} places the $G$ inputs of one output across DRAM columns and computes the $N_{out}$ outputs \emph{sequentially}.
A conventional compiler auto-vectorizer~\cite{AutoVect65:online, sarda2015llvm} produces this mapping when it vectorizes the innermost loop that combines the $G$ inputs. This mapping requires $N_{out}\cdot\lceil G/W\rceil$ column-parallel operations and leaves DRAM columns unused when $G<W$.
The \emph{output-stationary mapping} instead places different output elements across DRAM columns.
At each of the $G$ steps, every DRAM column updates one output using its next input element.
\prop can identify this mapping from the semantics of the data-parallel pattern. The mapping requires $G\cdot\lceil N_{out}/W\rceil$ column-parallel operations and exposes more column-level parallelism when $N_{out}>G$.
For each \texttt{stage}, \prop computes the operation counts of both mappings and selects the mapping with the smaller count.
When both $N_{out}$ and $G$ are \emph{smaller} than the subarray width $W$, \prop selects the output-stationary mapping when  $N_{out}\ge G$.
The output-stationary mapping benefits stages with many independent outputs, such as the matrix kernels implemented with \texttt{group}.
In contrast, the reduction-in-lane mapping benefits stages with few outputs, such as \texttt{reduce}.}

\paratitle{\omcrii{\underline{T3}: Bulk-Bitwise Operation Translation}} \gficcd{\omcrii{This} transformation translates each \texttt{stage} into a schedule of SIMDRAM bulk-bitwise operations, such as addition and multiplication.
SIMDRAM implements each bulk-bitwise operation as a precompiled sequence of majority and NOT operations executed bit-serially across the subarray columns.
\gficcdii{\omcri{The MIMDRAM compiler~\cite{mimdramextended,mimdramgit} compiles the generated schedule into the final \gls{PIM} binary.}}}


\subsection{\omcri{Discussion \& Limitations}}

\paratitle{\gficcd{\omcri{Backend} Selection}} \gficcd{\prop currently selects the target \omcri{backend} \emph{manually} at build time, i.e., the programmer chooses whether a \texttt{Pipeline} compiles for \gls{PnM} or \gls{PuM}.
Because the \omcri{frontend} program is identical for both substrates, a future system could defer this choice to runtime by emitting a \emph{fat binary} that contains both lowerings and dispatching each kernel to the substrate that best matches its characteristics, or even partition a single application across co-located \gls{PnM} and \gls{PuM} substrates for cooperative execution.
We leave such runtime-informed \omcri{backend} selection to future work.}

\paratitle{\gficcd{Adding New \omcri{Backend}s}} \gficcd{Supporting a new substrate requires implementing a new \omcri{backend}, i.e., a code template for the substrate together with the transformations that populate it from the \texttt{Pipeline}'s global stage and argument lists.
The \omcri{frontend} (\omcri{i.e.,} the data-parallel pattern APIs and the dataflow programming interface) and the substrate-independent analyses are reused unchanged \omcri{across all backends}, so the added effort is confined to the substrate-specific organization.}

\paratitle{\gficcd{\omcri{Backend} Generalizability}} \gficcd{The two \omcri{backend}s \omcri{(\gls{PnM} and \gls{PuM})} generalize beyond the specific substrates we target.
The \gls{PnM} \omcri{backend} applies to other near-bank architectures \omcri{(e.g.,~\cite{gao2017tetris, cho2020mcdram, shin2018mcdram, gu2020ipim, skhynixpim, kwon202125,yazdanbakhsh2018dram})} that expose a non-unified memory space with explicit host-to-core data transfers.
The \gls{PuM} \omcri{backend} is more broadly portable, since many bit-serial \gls{PuM} substrates, whether built on DRAM~\omcri{\cite{xin2020elp2im,oliveira2025proteus,li2017drisa, oliveira2026pipedram, deng2018dracc, li2018scope, shin2024processing}}, NVM~\gficcdii{\cite{truong2021racer, li2016pinatubo,fujiki2018memory,zha2020hyper}}, or SRAM~\gficcdii{\cite{dualitycache, eckert2018neural, aga2017compute,wang2019bit}}, share the same essential organization, i.e., a wide array of columns that compute a common operation bit-serially \omcri{using a} bit-vertical layout.
Retargeting the \gls{PuM} \omcri{backend} to such a substrate primarily requires adjusting the column width and the cost model.}

\paratitle{\gficcd{Limitations}} \gficcd{\prop's patterns operate on one-dimensional vectors, so multi-dimensional data must be flattened by the programmer. \gficcdiii{The \texttt{group} pattern recovers the structure that flattening removes, since it partitions the flattened vector into fixed-size groups that the pattern body treats as rows: our \texttt{GEMV} and \texttt{MLP} pipelines express matrix operands this way, and stencil-shaped kernels compose \texttt{window} with \texttt{group}. Kernels whose access pattern is neither row-wise nor windowed, such as matrix transposition, do \emph{not} map onto this interface.}
In addition, the \gls{PuM} \omcri{backend} currently lowers the data-independent patterns (\texttt{map}, \texttt{reduce}, and \texttt{group}), whereas the data-dependent patterns \texttt{filter} and \texttt{scan}, whose control flow maps poorly onto bit-serial execution, remain future work for \gls{PuM}.} \gficcdiii{\gficcd{Finally, our performance results \omcri{(\cref{sec:eval})} come from simulation rather than from a real \gls{PIM} system. Every system we evaluate runs under the \emph{same} simulator and the \emph{same} transfer model, so the comparison across systems, on which our conclusions rest, is \omcri{fair}; absolute execution times on real hardware will differ.}}

\section{\omcri{Experimental} Methodology}
\label{sec:methodology}

\subsection{\gficcd{Systems and Baselines}}
\gficcd{We evaluate \prop on two general-purpose \gls{PIM} substrates. We implement \prop in C++ as a shared library that can be distributed independently of the application.}

\paratitle{\omcri{\gls{PnM} Analysis}} {For \gls{PnM}, we target a UPMEM system~\omcri{\cite{gomez2022benchmarking, upmem2018, gomez2021benchmarkingcut, gomez2023evaluating}} with a 2-socket Intel Xeon Silver 4110 host CPU~\omcri{\cite{4110}} at \SI{2.10}{\giga\hertz}, \SI{128}{\giga\byte} of DDR4-2400, and 20 UPMEM PIM DIMMs that provide \SI{160}{\giga\byte} of PIM-capable memory across 2,560 PIM cores~\cite{upmem}. We obtain UPMEM performance from the cycle-accurate uPIMulator~\cite{hyun2024pathfinding}, which its authors validate against real silicon with 98.4\% single-PIM-core accuracy \omcri{and 83.6\% multi-PIM-core accuracy. We} use 11 PIM threads per core to fill the PIM-core instruction pipeline~\cite{gomez2022benchmarking}. \gficcdiii{We model CPU--PIM transfer time analytically, with bandwidth and per-transfer latency coefficients calibrated against published characterization of the commercial UPMEM system~\cite{gomez2022benchmarking}.
We report end-to-end \omcri{execution} time rather than PIM-core time alone because CPU--PIM transfers dominate many \gls{PnM} workloads~\omcri{\cite{lee2024pimmmu, gomez2022benchmarking}}.}}

\paratitle{\omcri{\gls{PuM} Analysis}} {For \gls{PuM}, we target SIMDRAM~\cite{hajinazarsimdram} and report its bit-serial bulk-bitwise execution time with its gem5-based cost model~\cite{mimdramgit} parameterized to a datapath of 65,536 DRAM columns per subarray. }

\paratitle{\omcri{Baselines}} \gficcd{We compare \prop against \gficcdiii{four} baselines:
\li~on \gls{PnM}, the hand-tuned PrIM workloads~\cite{gomez2022benchmarking}, and \lii~SimplePIM~\cite{chen2023simplepim}, a prior high-level UPMEM programming framework that still requires the user to manage CPU--PIM and PIM--PIM communication explicitly;
gficcdiii{\liii~on \gls{PnM}, CINM~\cite{khan2022cinm}, an MLIR-based \omcri{compiler targeting} both \gls{PIM} paradigms (\omcri{we build CINM from its released source code and use its UPMEM
  backend to compile each workload});}
\gficcdiii{\liv~on \gls{PuM}, MIMDRAM~\cite{mimdramextended}, the state-of-the-art \gls{PuM} compiler, which \emph{automatically} generates bit-serial instructions for the SIMDRAM~\cite{hajinazarsimdram} substrate by exploiting LLVM's loop auto-vectorization engine~\cite{mimdramgit}. Loop auto-vectorization is the code-generation mechanism of state-of-the-art \gls{PuM} systems~\omcri{\cite{mimdramextended, oliveira2025proteus, oliveira2026pipedram}}; our \gls{PuM} baseline is thus the strongest \emph{automatic} code generator available for this substrate.}}

\subsection{Workloads} 

\gficcd{For the \gls{PnM} substrate, we \omcri{use} workloads from the PrIM benchmark suite~\omcri{\cite{gomez2022benchmarking,gomezluna2021repo, gomez2021benchmarkingcut}} that exercise all six \omcri{of} \gficcdii{\prop's data-parallel} patterns, each expressed as a short \prop pipeline. Vector addition (\texttt{VA}) uses \texttt{map}, select (\texttt{SEL}) uses \texttt{filter}, unique (\texttt{UNI}) uses \texttt{window+filter} of size two, time-series similarity (\texttt{TS}) uses \texttt{window}, reduction (\texttt{RED}) and image histogram (\texttt{HST-S}, \texttt{HST-L}) use \texttt{reduce}, prefix sum (\texttt{SCAN}) uses \texttt{scan}, and matrix-vector multiplication (\texttt{GEMV}) and a three-layer multilayer perceptron (\texttt{MLP}) use \texttt{group}.}
\gficcd{To compare \prop against \omcri{the SimplePIM} high-level programming framework~\omcri{\cite{chen2023simplepim}}, we \omcri{also} run \texttt{VA}, \texttt{RED}, and image histogram, together with three machine-learning kernels, namely linear regression, logistic regression, and k-means clustering, each expressed as a \texttt{reduce} pipeline.}

\gficcd{For the \gls{PuM} substrate, we use twelve data-parallel kernels previously used to evaluate \gls{PuM} substrates~\omcri{\cite{mimdramextended, oliveira2025proteus, oliveira2026pipedram}}, which are drawn from benchmark suites such as PolyBench~\cite{pouchet2012polybench}, Rodinia~\cite{che_iiswc2009}, PARSEC~\cite{bienia2008parsec}, and Phoenix~\cite{yoo_iiswc2009}, each expressed with \prop's patterns and run at its standard problem size. 
The nested-loop kernels \texttt{2mm}, \texttt{3mm}, gemm (\texttt{gmm}), covariance (\texttt{cov}), \texttt{pca}, doitgen (\texttt{dg}), kmeans (\texttt{km}), gramschmidt (\texttt{gs}), heartwall (\texttt{hw}), and backprop (\texttt{bp}) use \texttt{group}, whereas the stencil kernel fdtd-apml (\texttt{fdtd}) and the motion-compensation kernel \texttt{x264} use \texttt{map}. }

\section{Evaluation}
\label{sec:eval}

\subsection{\omcri{Programming Effort} Improvement}

\gficcd{We first quantify how much \prop reduces programming effort on the \gls{PnM} substrate, using lines-of-code (LOC) as a productivity metric~\cite{boehm1984software}.\footnote{\gficcd{We report LOC only for the \gls{PnM} substrate. On \gls{PuM}, the MIMDRAM compiler~\cite{mimdramextended, mimdramgit} automatically generates bit-serial microcode through its \omcri{backend}, so the user writes \emph{no} substrate-specific code.}} We compare \prop against the hand-tuned PrIM workloads~\cite{gomez2022benchmarking} and against SimplePIM~\cite{chen2023simplepim}. 
We count only \emph{effective} UPMEM-programming code and exclude file loading, host memory allocation, variable definition, and timing as follows: 
\li~for PrIM, we count the hand-written PIM kernel each workload requires, 
\lii~for SimplePIM, the user-provided map and reduce callbacks together with the communication and table-management calls, and 
\liii~for \prop\omcri{,} the pipeline specification, namely the pattern stages and their data declarations.\footnote{\omcri{The reported LOC values measure workload-specific PIM-programming code rather than total application or framework size. A small LOC count does \emph{not} imply a small computation, since source-code length is \emph{independent} of input size, and each comparison implements the same computation at the same problem size.}} }
\gficcd{Table~\ref{tab:loc} reports the result for our 13 workloads. 
\omcri{Of the 13 workloads, seven have only a hand-tuned PrIM implementation, three have both hand-tuned PrIM and SimplePIM implementations, and three (linear regression, logistic regression, and k-means) have only a SimplePIM implementation. We make two observations.}}

\begin{table}[ht]
    \centering
    \caption{Lines-of-code (LOC) comparison.}
    \label{tab:loc}
    \tempcommand{0.8}
    \resizebox{\linewidth}{!}{
\begin{tabular}{@{}l||lll|l|l@{}}
\toprule
\textbf{Workload} & \textbf{\begin{tabular}[c]{@{}l@{}}LOC \\ PrIM~\cite{gomez2022benchmarking}\end{tabular}} & \textbf{\begin{tabular}[c]{@{}l@{}}LOC \\ SimplePIM~\cite{chen2023simplepim}\end{tabular}} & \textbf{\begin{tabular}[c]{@{}l@{}}LOC \\ \prop\end{tabular}} & \textbf{\begin{tabular}[c]{@{}l@{}}LOC Red.\\ (vs. PrIM)\end{tabular}} & \textbf{\begin{tabular}[c]{@{}l@{}}LOC \omcri{Reduction} \\ (vs. SimplePIM)\end{tabular}} \\ \midrule
\texttt{VA} & 78 & 14 & 6 & 92\% & 57\% \\
\texttt{SEL} & 123 & - & 4 & 97\% & - \\
\texttt{UNI} & 147 & - & 6 & 96\% & - \\
\texttt{RED} & 153 & 14 & 5 & 97\% & 64\% \\
\texttt{GEMV} & 198 & - & 11 & 94\% & - \\
\texttt{HST-S} & 116 & 21 & 8 & 93\% & 62\% \\
\texttt{SCAN} & 192 & - & 4 & 98\% & - \\
\texttt{HST-L} & 142 & - & 8 & 94\% & - \\
\texttt{TS} & 155 & - & 10 & 94\% & - \\
\texttt{MLP} & 171 & - & 12 & 93\% & - \\
\texttt{LINREG} & - & 43 & 15 & - & 65\% \\
\texttt{LOGREG} & - & 58 & 23 & - & 60\% \\
\texttt{KMEANS} & - & 61 & 22 & - & 64\% \\ \midrule
\textbf{GMean} & \textbf{143} & \textbf{29} & \textbf{7} & \textbf{95\%} & \textbf{62\%} \\ \bottomrule
\end{tabular}
}
\end{table}

{First, across the 10 workloads with a PrIM baseline, \prop reduces LOC by up to 98\% (95\% on average), because it raises the abstraction level so the user writes \emph{only} the target kernel rather than the manual DRAM copy streaming, handshaking between \gls{PIM} threads, and data transfer code that PrIM workloads manually \omcri{require}. Second, across the six workloads SimplePIM supports, \prop further reduces LOC by up to 65\% (62\% on average), since the user writes neither the data-communication nor the metadata-bookkeeping code that SimplePIM still requires. 
We conclude that \prop substantially eases \gls{PnM} \omcri{programming effort}.}

\subsection{Performance Analysis}

\begin{figure}[ht]
    \centering
    \includegraphics[width=\linewidth]{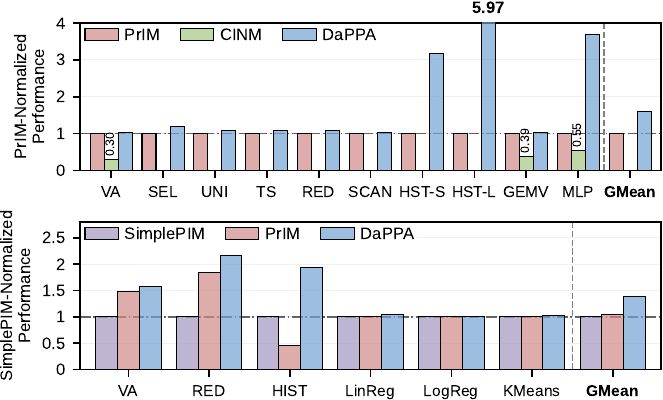}
    \caption{\omcri{\gficcd{\prop performance on the UPMEM substrate \omcrii{versus} hand-tuned PrIM~\cite{gomez2022benchmarking} (top) and SimplePIM~\cite{chen2023simplepim} (bottom).} \gficcdiii{The top panel also reports CINM~\cite{khan2022cinm} on the three workloads it can deploy.}}}
    \label{fig:eval_pnm}
\end{figure}

\paratitle{\omcri{\gls{PnM} Performance}}
\gficcd{Fig.~\ref{fig:eval_pnm} shows \prop's \gls{PnM} performance on the UPMEM substrate. 
The top \omcri{plot} reports end-to-end execution time across ten workloads normalized to their hand-tuned PrIM implementations, and the bottom \omcri{plot} compares \prop against both PrIM~\cite{gomez2022benchmarking} and the SimplePIM framework~\omcri{\cite{chen2023simplepim}} on the six workloads SimplePIM supports. 
We make \gficcdiii{five} observations from the figure.}
\gficcd{First, on average across all ten \gficcdii{PrIM} workloads \gficcdii{(Fig.~\ref{fig:eval_pnm}, top)}, \prop matches or exceeds the performance of hand-tuned PrIM implementations, \omcri{providing} 1.6$\times$ \omcri{higher} end-to-end performance (up to 6.0$\times$ for \texttt{HST-L}).}
\gficcd{Second, \prop's largest end-to-end performance gains \omcri{against hand-tuned PrIM workloads} (i.e., \texttt{HST-S} at 3.2$\times$, \texttt{HST-L} at 6.0$\times$, and \texttt{MLP} at 3.7$\times$ that of PrIM), come from a \emph{leaner} and \emph{optimized} code for the PIM kernels, which is more visible in the histogram kernels. 
To map an element to a bin, the histogram kernel multiplies the element by the histogram's bin count. 
PrIM passes that bin count to the kernel as a runtime argument, so the multiply \omcri{operation requires a} true integer \omcri{multiplication}. Since UPMEM PIM cores provide hardware support to only 8-bit multiplication~\cite{gomez2022benchmarking}, \omcri{the UPMEM compiler} expands \omcri{a general 32-bit integer multiplication} into multi-cycle microcode that dominates the execution time. 
\prop instead fixes the bin count as a compile-time constant in the kernel it generates, so the compiler replaces the multiply with a constant shift-and-add (a single shift when the bin count is a power of two), \omcri{thereby} removing the need for the multiply operation. 
\texttt{HST-L} widens the gap further because PrIM serializes its histogram updates behind a mutex, whereas \prop replicates the accumulator per \gls{PIM} thread and updates it lock-free. \gficcdiii{\prop applies both optimizations \emph{automatically} while generating the kernel, from the \emph{same} user-written pipeline: the programmer neither identifies nor implements them. \prop's performance gains are therefore \emph{not} separable from its code generation.}}
\gficcd{Third, the remaining seven workloads (\texttt{VA}, \texttt{SEL}, \texttt{UNI}, \texttt{TS}, \texttt{RED}, \texttt{SCAN}, and \texttt{GEMV}) stay at parity with PrIM because their end-to-end \omcri{execution} time is dominated by CPU--PIM data transfer in both direction. 
}
\gficcdiii{Fourth, Fig.~\ref{fig:eval_pnm} (top) compares \prop against CINM~\cite{khan2022cinm}, the closest prior framework, which also spans both \gls{PIM} paradigms. 
On the three workloads it supports, CINM reaches 0.30$\times$, 0.39$\times$, and 0.55$\times$ the performance of hand-tuned PrIM, whereas \prop matches or exceeds PrIM on all three.\footnote{\omcri{CINM successfully deploys \emph{only} three of our ten workloads (\texttt{VA}, \texttt{GEMV}, and \texttt{MLP}), since its fixed linear algebra operator catalog \emph{cannot} express the filtering, histogram, and search workloads, and its released compiler \emph{cannot} \omcrii{generate} UPMEM \omcrii{code for} the reduction and scan operators.}} 
CINM's performance gap comes from its execution model, since it re-launches a fixed-tile kernel once per tile (16 launches for \texttt{GEMV} and 12 for \texttt{MLP} on each PIM core), re-broadcasting operands and paying the per-launch transfer latency every time, whereas \prop generates one specialized kernel for the whole input.}

\gficcd{Fifth, \prop is 1.4$\times$ faster than SimplePIM, on average \gficcdii{across the six SimplePIM workloads} \omcri{(as shown in \gficcdii{Fig.~\ref{fig:eval_pnm}, bottom)}}. 
\omcri{On \texttt{VA}, \prop and PrIM achieve 1.6$\times$ and 1.5$\times$ the performance of SimplePIM, respectively. On \texttt{RED}, \prop \omcrii{provides} 2.2$\times$ the performance of SimplePIM, whereas PrIM \omcrii{provides} 1.8$\times$ the performance of SimplePIM. SimplePIM calls the reduction function through a function pointer for every input element, adding overhead to each reduction step. \prop eliminates this overhead by inlining the user-supplied reduction function into the generated kernel. \prop also keeps the running reduction result (the accumulator) in a register while processing each block and writes it to the PIM core's scratchpad memory only once per block. This avoids reading and writing the accumulator in the PIM core's scratchpad memory for every input element.}
On the histogram \gficcdii{kernel}, \prop is 1.9$\times$ faster than SimplePIM, and PrIM is in fact 2.2$\times$ \emph{slower} than SimplePIM for the runtime-multiply reason above. 
On the histogram \gficcdii{kernel}, \prop is 1.9$\times$ faster than SimplePIM, and PrIM is 2.2$\times$ \emph{slower} than SimplePIM \omcri{because PrIM computes the histogram bin index using an integer multiplication for every input element, which is costly due to the lack of hardware support for 32-bit integer multiplication in UPMEM PIM cores~\cite{upmem, gomez2022benchmarking, gomez2021benchmarkingcut}.}
On the three machine-learning reductions \texttt{LinReg}, \texttt{LogReg}, and \texttt{KMeans}, \prop matches SimplePIM, \omcri{showing that \prop automatically generates reduction kernels that match the performance of SimplePIM's hand-written kernels.}}
\omcri{\gficcd{We conclude that \prop matches or exceeds the performance of hand-tuned PrIM implementations and the SimplePIM framework while requiring only a high-level data-parallel program. \prop \omcrii{provides} its end-to-end performance gains by automatically generating specialized PIM kernels that reduce computation and synchronization overheads.}}

\begin{figure}[ht]
    \centering
    \includegraphics[width=\linewidth]{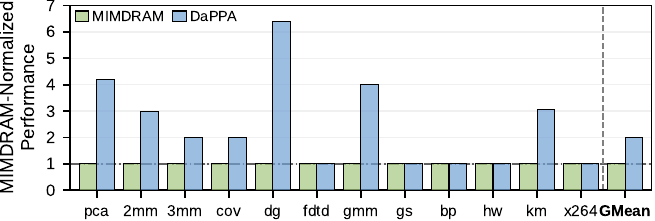}
    \caption{\omcri{\prop performance on  SIMDRAM~\cite{hajinazarsimdram} \gls{PuM}\gficcdiii{, normalized to the code generated by the state-of-the-art MIMDRAM compiler~\cite{mimdramextended}}.}}
    \label{fig:eval_simdram}
\end{figure}

\paratitle{\omcri{\gls{PuM} Performance}}
\gficcd{Fig.~\ref{fig:eval_simdram} shows \prop's performance on the SIMDRAM substrate for twelve \gls{PuM} kernels \omcri{(\cref{sec:methodology})}, \gficcdiii{normalized to the code that the state-of-the-art MIMDRAM compiler~\cite{mimdramextended, mimdramgit} automatically generates for the same SIMDRAM~\cite{hajinazarsimdram} substrate.} We make three observations from the figure.}
\gficcd{First, on average across all twelve kernels, \prop provides 2.0$\times$ the performance of \gficcdiii{MIMDRAM} code (up to 6.4$\times$ for \texttt{dg}).}

\gficcd{\omcri{Second, \prop's performance gains mainly come from mapping independent outputs to different DRAM columns of a subarray. 
Each subarray provides 65{,}536 DRAM columns that can compute in parallel. 
\omcrii{The} MIMDRAM compiler exploits this parallelism \emph{only} across iterations of the \emph{innermost} loop. 
In matrix multiplication, these iterations compute the products that are summed to produce a single output element. 
As a result, MIMDRAM assigns the products for one output element to different DRAM columns, while computing successive output elements sequentially. 
In our seven matrix-multiplication-based kernels (\texttt{2mm}, \texttt{3mm}, \texttt{gmm}, \texttt{cov}, \texttt{pca}, \texttt{dg}, and \texttt{km}), each output requires summing only 10{,}000 to 32{,}000 products. 
Therefore, this mapping \omcrii{done by the MIMDRAM compiler} leaves 51\% to 85\% of the subarray's DRAM columns idle during computation. 
To use more of \omcrii{the available} parallelism, \prop's backend assigns a different output element to each DRAM column, allowing many outputs to be computed \emph{simultaneously}. 
Each DRAM column retains its partial sum and updates it as successive input values arrive, following \prop's output-stationary mapping (\cref{sec:impl:pum}) Across these seven kernels, \prop provides 3.3$\times$ the performance of code generated by the MIMDRAM compiler, on average.}}
\gficcd{Third, for the remaining five kernels, \prop matches \gficcdiii{MIMDRAM \omcri{compiler code's}} performance. 
These kernels are either element-wise maps and stencils with no reduction to tile (\texttt{fdtd} and \texttt{x264}) or kernels whose output space does \emph{not} exceed their reduction (\texttt{gs}, \texttt{hw}, and \texttt{bp}), so \prop's \omcri{backend} selects the same inner-loop vectorization \gficcdiii{MIMDRAM's compiler} uses and emits identical microcode.}
\gficcd{\omcri{We conclude that \prop targets the \gls{PuM} substrate \emph{without} any change to the user program and matches or exceeds the performance of code generated by the MIMDRAM compiler on every evaluated kernel.}}

\section{Related Work}

\omcri{To our knowledge, \prop is the first data-parallel pattern-based programming framework that \emph{automatically} generates substrate-specific computation and data organization for both \gls{PnM} and \gls{PuM} substrates from the \emph{same} user program. We highlight \prop's key contributions by contrasting it with prior PIM programming frameworks and compilers.}

\paratitle{Programming Frameworks for \gls{PnM}} \omcri{Prior works ease \gls{PnM} programming \omcrii{effort} through language extensions, parallel patterns, and optimized libraries~\cite{fraguela2003programming,hall1999mapping,jacob2016compiling,hong2024programming,chen2023simplepim}. These interfaces help programmers assign computation to PIM cores and manage its data. Specialized libraries further simplify arithmetic, sparse linear algebra, and graph workloads~\omcrii{\cite{item2023transpimlib,giannoula2022sparsep,nai2017graphpim,giannoula2024pygim, rhyner2024pim}}. \prop provides composable patterns and automatically derives communication from computation, while supporting both PIM paradigms.}

\paratitle{Programming Frameworks for \gls{PuM}} \omcri{Prior works enable programmers to express computation within memory arrays using tensor operations, GPU programming models, dataflow graphs, computational skeletons, or explicit instructions~\cite{fujiki2018memory,dualitycache,leitersdorf2024pypim,peng2023chopper,wang2021ssr,truong2026mpu}. 
These interfaces simplify programming for their target \gls{PuM} systems. 
\omcrii{In contrast,    } \prop exposes a \emph{shared} pattern interface that can target either \gls{PnM} or \gls{PuM} \omcrii{systems} without changing the user program.}

\paratitle{Programming Frameworks for \gls{PnM} and \gls{PuM}} \omcri{Prior frameworks span both \gls{PnM} and \gls{PuM} paradigms via set operations, C-based dataflow programs, or high-level operators~\cite{besta2021sisa,wang2023infinity,khan2022cinm,hadidi2017cairo,ahmed2019compiler,hwang2025pim}. 
They enable computation across near-memory processors and memory arrays, establishing programming support for heterogeneous PIM systems. 
\prop differs by exposing composable data-parallel patterns with user-defined computation and automatically generating the execution and data organization required by each backend.}

\paratitle{Compiler Support for \gls{PnM}} \omcri{Prior \omcrii{\gls{PnM}} compilers identify suitable computation for near-memory execution or translate programmer annotations into accelerator code~\cite{solihin2001automatic,hsieh2016transparent,hadidi2017cairo,ahmed2019compiler,jacob2016compiling,hong2024programming}. 
Tensor compilers also optimize data placement and execution schedules, including joint host/PIM code generation~\cite{shin2023pimflow,shin2025atim,xie2025unindp,yang2026dcc}. 
For example, \omcri{DCC~\cite{yang2026dcc} is a concurrent work that targets machine-learning applications and comprises code and data optimizations, but targets \emph{only} \gls{PnM} substrates.}
\prop derives computation and communication from explicit pattern semantics and supports both \omcrii{\gls{PnM} and \gls{PuM}} paradigms.}

\paratitle{Compiler Support for \gls{PuM}} \omcri{Prior \gls{PuM} compilers translate dataflow programs, computational skeletons, and C/C++ loops into memory-array operations~\cite{peng2023chopper,wang2021ssr,zha2020hyper,mimdramextended, dualitycache, fujiki2018memory}. Other \omcrii{compiler-based approaches~\cite{vadivel2020tdo,qu2024cim,zhao2025cim,tang2024pimlc}} recognize matrix operations or compile neural-network graphs while optimizing data movement and array utilization. 
\prop complements these work by deriving computation and \omcrii{data} mapping from data-parallel patterns, \omcrii{and} then using the MIMDRAM compiler to generate the final \gls{PuM} binary.}

\paratitle{Compiler Support for \gls{PnM} and \gls{PuM}} \omcri{\omcrii{A handful of prior works support} both \gls{PnM} and \gls{PuM} via hierarchical compiler representations, runtime compilation, and optimized computation mapping~\cite{khan2022cinm,wang2023infinity,liu2025optipim}. These approaches translate supported computation into device-specific execution or optimize its placement across PIM resources. In contrast, \prop generates kernels and data organization from the programmer's composition of data-parallel patterns. \omcrii{Prior works'~\cite{khan2022cinm,wang2023infinity,liu2025optipim}} data mapping and scheduling techniques are complementary to \prop's pattern-based code generation.}
\section{Conclusion}
\label{sec:conclusion}

\gficcd{\omcri{We introduce \prop, a high-level data-parallel programming framework that provides a single programming interface for both \gls{PnM} and \gls{PuM} substrates.
The key idea of \prop is to decouple the expression of computation from substrate-specific data organization.
Programmers express computation once as a pipeline of data-parallel patterns, while specialized backends automatically generate the computation and data organization required by each substrate.
Our evaluation shows that 
\omcrii{\li~for \gls{PnM} systems,} \prop matches or exceeds the performance of hand-tuned PrIM implementations and the SimplePIM framework on UPMEM while substantially reducing programming effort\omcrii{;
\lii~for \gls{PuM} systems,} \prop outperforms code generated by the MIMDRAM compiler;
\omcrii{\liii~for a combined \gls{PnM} and \gls{PuM} system,} \prop outperforms CINM on every workload supported by both frameworks.} We believe and hope that \prop \omcri{will ease the} adoption of current and future general-purpose \gls{PIM} systems.
\omcri{To \omcrii{enable} further research on \gls{PIM} programmability, we freely open-source \prop at \url{https://github.com/CMU-SAFARI/DaPPA}.}}

\section*{\gfcr{Acknowledgments}}

\gfcr{We thank the anonymous reviewers of \gfcr{\omcrii{PACT SRC 2023,} ICS 2025, PACT 2025, HPCA 2026, ISPASS 2026, ICCD 2026} for their feedback. 
We thank the SAFARI Research Group members for providing a stimulating intellectual \omcri{and scientific} environment. 
We acknowledge the generous gifts from our industrial partners, including Google, Huawei, Intel, Microsoft\omcri{, and VMware}. This work is supported in part by the ETH Future Computing Laboratory (EFCL), Semiconductor Research Corporation, AI Chip Center for Emerging Smart Systems (ACCESS), and European Union's Horizon programme for research and innovation [101047160 - BioPIM].}

{
  \bstctlcite{IEEEexample:BSTcontrol}
  \let\OLDthebibliography\thebibliography
  \renewcommand\thebibliography[1]{
    \OLDthebibliography{#1}
    \setlength{\parskip}{0pt}
    \setlength{\itemsep}{0pt}
  }
  \bibliographystyle{IEEEtran}
  \bibliography{refs}
}

\end{document}